\documentclass[final,12pt]{elsarticle}
\usepackage[T1]{fontenc}
\usepackage[utf8]{inputenc}

\usepackage{amsmath,amsthm,amsfonts,amssymb,amscd}
\usepackage{bm}
\usepackage{mathtools}
\usepackage{mathrsfs}

\usepackage{graphicx}
\usepackage{float}
\usepackage{multirow,booktabs}
\usepackage{subcaption}
\usepackage{caption}

\usepackage{algorithm}
\usepackage{algpseudocode}

\usepackage{hyperref}

\usepackage{tikz}
\usetikzlibrary{arrows.meta, positioning, fit}

\usepackage[lighttt]{lmodern}
\makeatletter
\algrenewcommand\ALG@beginalgorithmic{\ttfamily}
\makeatother

\newlength\myindent
\makeatletter
\def\ps@pprintTitle{%
  \let\@oddhead\@empty
  \let\@evenhead\@empty
  \let\@oddfoot\@empty
  \let\@evenfoot\@oddfoot
}
\makeatother

\begin{document}

\begin{frontmatter}
    \title{Uncertainty calibration for latent-variable regression models}
    
    \author[1]{Zina-Sabrina {Duma}\corref{cor1}} 
    \cortext[cor1]{Corresponding author}
    \ead{Zina-Sabrina.Duma@lut.fi}
    \author[2]{Otto {Lamminpää}}
    \author[2]{Jouni {Susiluoto}}
    \author[1]{Heikki {Haario}}
    \author[3]{Ting {Zheng}}
    \author[1]{Tuomas {Sihvonen}}
    \author[2]{Amy {Braverman}}
    \author[3]{Philip A. {Townsend}}
    \author[1]{Satu-Pia {Reinikainen}}
    \address[1]{LUT University, Yliopistonkatu 34, Lappeenranta 53850 Finland}
    \address[2]{Jet Propulsion Laboratory, California Institute of Technology, 4800 Oak Grove Drive, La Cañada Flintridge, CA 91011, USA}
    \address[3]{Department of Forest and Wildlife Ecology, University of Wisconsin-Madison, 1630 Linden Drive, Madison, WI 53706, USA}
    
    \begin{abstract}
    Uncertainty quantification is essential for scientific analysis, as it allows for the evaluation and interpretation of variability and reliability in complex systems and datasets. In their original form, multivariate statistical regression models (partial least-squares regression, PLS, principal component regression, PCR) along with their kernelized versions (kernel partial least-squares regression, K-PLS, kernel principal component regression, K-PCR), do not incorporate uncertainty quantification as part of their output. In this study, we propose a method inspired by conformal inference to estimate and calibrate the uncertainty of multivariate statistical models. The result of this method is a point prediction accompanied by prediction intervals that depend on the input data. We tested the proposed method on both traditional and kernelized versions of PLS and PCR. The method is demonstrated using synthetic data, as well as laboratory near-infrared (NIR) and airborne hyperspectral regression models for estimating functional plant traits. The model was able to successfully identify the uncertain regions in the simulated data and match the magnitude of the uncertainty. In real-case scenarios, the optimised model was not overconfident nor underconfident when estimating from test data: for example, for a 95\% prediction interval, 95\% of the true observations were inside the prediction interval.
    \end{abstract}
    
\end{frontmatter}

\section{Introduction}
Uncertainty quantification (UQ) is essential for evaluating the reliability of a regression model. In the context of regression, uncertainty can arise from multiple sources, including uncertainty in model parameters, uncertainty in fitted values for calibration data, and uncertainty in predictions for new samples. Decisions made based on uncertain information can have significant consequences \cite{walker2003defining}. UQ enhances risk management and helps identify the weaknesses of a model by highlighting areas of low certainty or where the model requires additional training data \cite{abdar2021review}. Continuous monitoring of uncertainty can also reveal model degradation over time or in response to changes in data \cite{jorgensen2023extensible}. In this work, we explicitly target predictive uncertainty for new samples, i.e., the uncertainty associated with future observations conditional on a trained regression model.

Latent variable (LV)–based methods, including principal component regression (PCR) and partial least-squares (PLS) regression, along with their kernelized extensions—kernel principal component regression (K-PCR) and kernel partial least-squares (K-PLS) regression—are widely used tools for spectral regression modeling \cite{passos2022tutorial}. Linear LV methods such as PCR and PLS have been used for many years and are particularly preferred in situations where (a) the predictor variables are highly collinear, (b) the number of observations is limited, and (c) model interpretability is an important consideration \cite{gromski2015tutorial}. As a result, PCR and PLS remain standard choices in applications such as instrument calibration \cite{wang2022recent}, remote sensing for plant trait regression \cite{kamoske2021leaf}, and process monitoring \cite{ji2022review}. While PCR and PLS model linear relationships between predictors and responses, their kernelized extensions were developed to capture nonlinear dependencies by implicitly mapping the data into a high-dimensional feature space. Kernel-based extensions are employed when nonlinear relationships between predictors and responses cannot be adequately captured by linear LV models.

Multivariate models traditionally return point estimates, without explicit quantification of predictive uncertainty for new observations \cite{zhang2009comparison}, even though some studies have utilized bootstrap analysis and jackknifing to quantify uncertainty \cite{faber1997propagation, martens2000modified, wentzell2015errors}. The original approach of bootstrapping involves resampling the input data matrices and re-computing the scores, loadings, and residuals \cite{babamoradi2013bootstrap}. Statistical analysis is then performed on the resampled outputs to assess variability in model predictions. Another approach is residual bootstrapping \cite{preisner2008uncertainty}, where the input measurements remain unchanged while the response variable is perturbed. In this case, the model residuals are treated as representative of predictive uncertainty; these residuals are randomly sampled and added back to the model estimates, resulting in a distribution of predicted values \cite{de2013discrimination, rocha2018classification}.

In this work, we focus specifically on predictive uncertainty for new samples, expressed in the form of prediction intervals (PIs) around point predictions. These intervals are intended to quantify the variability of future observations conditional on a fixed, trained regression model and the available calibration data. To this end, we propose a novel uncertainty calibration method for LV models inspired by conformal inference. Conformal inference, also known as conformal prediction, is a distribution-free statistical framework used in machine learning to construct prediction intervals with guaranteed marginal coverage \cite{gibbs2021adaptive}. This framework leverages historical data to determine uncertainty bounds for future predictions, allowing each prediction to be accompanied by an input-dependent prediction interval.

The original concept of conformal inference involves dividing the available data into a training set and an uncertainty calibration set \cite{einbinder2022training}. A regression model is fitted on the training data, and prediction residuals are evaluated on the calibration set. Empirical quantiles of these residuals are then used to construct prediction intervals for new samples, yielding marginal coverage guarantees under the assumption of exchangeability. In its standard form, conformal inference produces a single, global residual quantile that is applied uniformly to all new predictions. While this approach is simple and robust, it does not account for input-dependent or heteroscedastic uncertainty. Extensions that localize conformal prediction to regions of the input space exist, but in high-dimensional regression problems—such as spectroscopy—defining meaningful neighborhoods or partitions directly in the original predictor space becomes challenging due to the curse of dimensionality \cite{lei2018distribution}.

In this paper, we propose a LV-based uncertainty calibration approach that adapts conformal inference to multivariate regression models. Instead of localizing uncertainty in the original high-dimensional predictor space, uncertainty is conditioned on the projections produced by PCR, PLS, and their kernelized counterparts. These latent variables provide a low-dimensional representation that captures the dominant systematic structure in the data. The proposed method proceeds as follows: (a) latent-variable scores from the calibration set are discretized into intervals for each retained LV; (b) empirical residual quantiles are estimated separately within each LV interval using calibration residuals; and (c) for a new sample, the interval-specific quantiles corresponding to its latent projections are combined using weights proportional to the explained variance of each LV. The resulting prediction interval is therefore input-dependent and reflects empirical uncertainty patterns observed in similar regions of the latent space, while remaining computationally efficient and compatible with standard PCR, PLS, K-PCR, and K-PLS workflows.

We illustrate the proposed uncertainty calibration procedure through four case studies. One case study is semi-synthetic, with real satellite data used as predictors and a simulated response with known predictive uncertainty. The second case study presents a soil moisture regression model from hyperspectral soil data, modeled with K-PCR. The third case study uses the benchmark corn NIR dataset to estimate moisture content with PLS. The final case study addresses a complex plant trait regression problem using airborne hyperspectral data and K-PLS.

\section{Theoretical background}\label{sec:background}

\subsection{Usual prediction interval estimation for latent-variable regression models}
\label{ssec:uqLV}

In LV-based regression, PIs quantify uncertainty on predicted values for new samples, i.e., intervals intended to contain the future observation $y_*$ associated with a new observation $\mathbf{x}_*$ \cite{montgomery2012regression,martens1989multivariate}. PIs are commonly obtained using either (i) analytic variance-based approximations derived from modeling assumptions, or (ii) resampling-based approaches such as bootstrap or jackknife procedures \cite{martens1989multivariate,faber2002uncertainty,martens2000modified}.

A classical variance-based approach assumes an additive noise model and approximates the predictive distribution using an estimated residual variance together with the prediction variance induced by uncertainty in the fitted regression coefficients \cite{montgomery2012regression}.
Under Gaussian noise and standard linear-model assumptions, a typical PI can be written as
\begin{equation}
    \widehat{C}_{1-\alpha}(\mathbf{x}_*) =
    \left[\hat{y}_* \pm t_{1-\alpha/2,\nu}\,\widehat{\sigma}_{\mathrm{pred}}(\mathbf{x}_*)\right],
\end{equation}
where $t_{1-\alpha/2,\nu}$ is a Student-$t$ quantile and $\widehat{\sigma}_{\mathrm{pred}}(\mathbf{x}_*)$ combines an estimate of the residual variance with uncertainty propagated from the fitted model \cite{montgomery2012regression}.
In LV regression, this propagation is less straightforward than in ordinary least squares because the regression coefficients depend on a dimension-reduction step (scores/loadings), and additional approximations (or empirical variance estimators) are often required \cite{martens1989multivariate,faber2002uncertainty}.

Resampling approaches reduce reliance on parametric distributional assumptions by repeatedly refitting the model on resampled data and using the induced variability in predictions to construct PIs \cite{efron1993bootstrap,davison1997bootstrap}.
In a “naïve” (pairs) bootstrap, observations are resampled with replacement, the LV model is refit for each bootstrap replicate, and the empirical distribution of $\hat{y}_*(b)$ is used to obtain PI endpoints via quantiles \cite{efron1993bootstrap,davison1997bootstrap}.
Related resampling ideas have also been developed specifically for bilinear LV modeling, including jackknife-type estimators for parameter uncertainty in PLS/PCR \cite{martens2000modified,faber2002uncertainty}.

A particularly relevant variant for regression is residual bootstrapping (fixed design), in which the predictor matrix is kept fixed and the response is perturbed by resampling residuals \cite{efron1993bootstrap,wu1986jackknife}.
After fitting a model, residuals $\hat{\varepsilon}_i = y_i-\hat{f}(\mathbf{x}_i)$ are resampled and added back to fitted values to create pseudo-responses, and the resulting ensemble of refits (or perturbed predictions) yields an empirical predictive distribution for constructing PIs \cite{efron1993bootstrap,davison1997bootstrap,faber2002uncertainty}.
This strategy is intuitive because it explicitly reuses the empirically observed error structure to quantify predictive uncertainty \cite{faber2002uncertainty}.

Despite their widespread use, variance-based and bootstrap-based PIs generally rely on modeling assumptions (e.g., approximate normality, homoscedasticity) and/or on the representativeness of the resampling scheme \cite{montgomery2012regression,efron1993bootstrap,davison1997bootstrap}.
Moreover, these approaches do not typically provide finite-sample distribution-free marginal coverage guarantees under minimal assumptions \cite{shafer2008tutorial,lei2018distributionfree}.
These considerations motivate conformal prediction, which constructs PIs from empirical quantiles of calibration residuals and provides distribution-free marginal coverage under exchangeability \cite{vovk2005algorithmic,shafer2008tutorial,lei2018distributionfree}.

\subsection{Conformal inference}\label{ssec:confInf}

Conformal inference, often referred to as conformal prediction, is a distribution-free framework for constructing prediction intervals with a user-specified marginal coverage level, e.g., $1-\alpha = 0.95$ \cite{vovk2005algorithmic, shafer2008tutorial}. 
The key idea is to use past errors on a held-out calibration set to determine how wide prediction intervals should be for future samples \cite{lei2018distributionfree, romano2019conformalized}.

Consider a regression model $\hat{f}(\cdot)$ trained on a training set $\mathcal{D}_{\mathrm{tr}} = \{(\mathbf{x}_i, y_i)\}_{i=1}^{n_{\mathrm{tr}}}$.
A separate calibration set $\mathcal{D}_{\mathrm{cal}} = \{(\mathbf{x}_j, y_j)\}_{j=1}^{n_{\mathrm{cal}}}$ is then used to quantify typical prediction errors through nonconformity scores \cite{vovk2005algorithmic}. 
A common choice for regression is the absolute residual \cite{lei2018distributionfree, angelopoulos2021gentle}:
\begin{equation}
    s_j = \big| y_j - \hat{f}(\mathbf{x}_j) \big|, \quad j=1,\ldots,n_{\mathrm{cal}}.
\end{equation}

Let $\widehat{q}_{1-\alpha}$ denote the empirical $(1-\alpha)$-quantile of $\{s_j\}_{j=1}^{n_{\mathrm{cal}}}$ \cite{shafer2008tutorial, angelopoulos2021gentle}. 
For a new input $\mathbf{x}_*$ with point prediction $\hat{y}_* = \hat{f}(\mathbf{x}_*)$, conformal prediction constructs the prediction interval \cite{lei2018distributionfree}
\begin{equation}
    \widehat{C}_{1-\alpha}(\mathbf{x}_*) =
    \big[\hat{y}_* - \widehat{q}_{1-\alpha},\; \hat{y}_* + \widehat{q}_{1-\alpha}\big].
\end{equation}

Under the standard exchangeability assumption (i.e., calibration samples and the new sample are identically distributed and order-invariant), this interval attains the target marginal coverage \cite{vovk2005algorithmic, shafer2008tutorial}
\begin{equation}
    \mathbb{P}\!\left( y_* \in \widehat{C}_{1-\alpha}(\mathbf{x}_*) \right) \ge 1-\alpha.
\end{equation}
A limitation of this basic ``global'' approach is that the same $\widehat{q}_{1-\alpha}$ is used for all test points, which may yield overly conservative intervals in low-uncertainty regions and under-responsive intervals in high-uncertainty regions \cite{lei2018distributionfree, romano2019conformalized}.

\begin{figure}[H]
\centering
\includegraphics[width=0.9\linewidth]{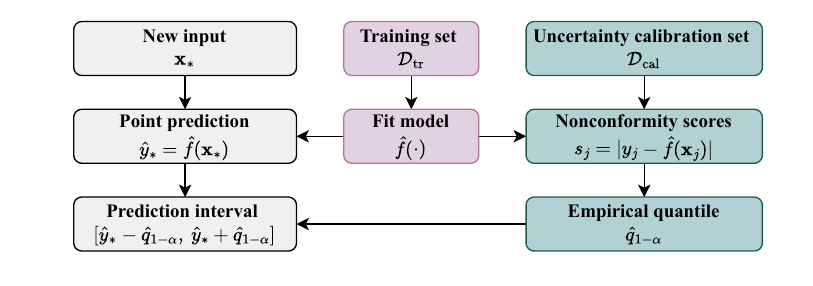}
\caption{Standard conformal inference workflow for regression. A model is fitted on a training set, calibration residuals are converted into nonconformity scores, and an empirical $(1-\alpha)$ quantile of these scores is used to transform a point prediction into a prediction interval for a new sample.}
\label{fig:conf_workflow}
\end{figure}

Figure~\ref{fig:conf_workflow} summarizes this standard conformal inference workflow and highlights the role of the calibration residual quantile in forming prediction intervals.

\subsection{Limitations of existing methods and proposed solutions}

To address input-dependent uncertainty, various localized conformal variants estimate different quantiles for different regions of the input space\cite{lei2018distributionfree}. In high-dimensional regression, however, defining such regions directly in the original predictor space can be challenging \cite{romano2019conformalized}. In this work, we instead localize uncertainty in the LV space produced by multivariate methods, enabling input-adaptive intervals while preserving the dimensionality reduction intrinsic to PCR/PLS and their kernelized counterparts.

From a conformal prediction perspective, the proposed approach can be interpreted as a localized split conformal method in which the conditioning variable is not the original predictor $\mathbf{x}$ but its latent-variable projection $\mathbf{t}$ produced by an LV regression model. The novelty lies in (i) defining the localization in the low-dimensional LV space intrinsic to PCR/PLS and their kernel extensions, which avoids neighborhood construction in the original high-dimensional predictor space, and (ii) aggregating LV-specific conditional residual quantiles using explained-variance weights to obtain a single PI half-width. This yields a conformal-inspired PI construction based on input-adaptive yet computationally lightweight LV models.

An important practical consideration is that not all sources of variability in multivariate models are informative for predictive uncertainty calibration. LVs that primarily capture non-systematic variation or measurement noise do not represent reproducible structure in the data and should therefore not be used to localize predictive uncertainty. Using such noise-dominated directions to calibrate PIs can lead to unstable or misleading uncertainty estimates. The proposed approach mitigates this issue by performing uncertainty localization in the LV space and weighting the contribution of each LV according to its explained variance. As a result, only systematic, data-supported variation influences the width of the PIs, while low-variance noise components have a limited effect. This contrasts with approaches that attempt to localize uncertainty directly in the original high-dimensional predictor space, where non-systematic variation can dominate distance or neighborhood-based measures.

This makes the proposed method particularly suitable for chemometric and spectroscopic models, where LVs are explicitly designed to separate systematic signal from unstructured noise. The proposed LV-localized calibration should therefore be interpreted as a conformal-inspired procedure: exact finite-sample marginal coverage is guaranteed in the global (single-interval) case, while localized calibration trades strict distribution-free guarantees for improved adaptivity to input-dependent uncertainty.

Split conformal prediction can be viewed as a residual-quantile calibration approach: it constructs a PI by estimating an empirical quantile of absolute prediction errors on held-out calibration samples. This is conceptually close to residual bootstrap methods that also reuse empirical residual behavior to quantify predictive uncertainty. A key difference is that conformal prediction uses a separate calibration set and, under exchangeability, provides a finite-sample marginal coverage guarantee without assuming Gaussian errors or a specific parametric noise model.

\section{Proposed mathematical methods}\label{sec:maths}

The proposed uncertainty calibration workflow presents common points for all the multivariate methods that involve latent projections: (i) the data is divided into three partions: the model training set, the uncertainty calibration set and the testing partiton; (ii) the model is trained on the training set; (iii) predictions are made with the model for the uncertainty calibration set; (iv) the latent variable projections of the uncertainty calibration set are divided into intervals; (v) the prediction residuals and residual quantiles are computed for each of the LV intervals; (vi) LV weights are calculated based on the explained variance of each LV.

When making new predictions with the model, the confidence predictions are evaluated as follows: (i) the new data point is projected onto the LVs; (ii) based on the LV projections, membership to LV intervals is assessed; (iii) the residual quantiles are weighted by the explained variance of the LV and summed up; (iv) the resulting residual quantiles are utilized to form the prediction intervals for the prediction.

The following section goes through the mathematics of this workflow, both for the uncertainty calibration procedure and making predictions with the model and assessing the uncertainty of the predictions with PCR. Extension to PLS, K-PLS and K-PCR is present in \textbf{Appendix B}.

\subsection{Data preparation}\label{ssec:data}

Let $\mathbf{X}$ be a predictor data matrix of size $n \times m$, which is divided into a training set, an uncertainty calibration set, and a testing set. The roles of these sets are defined as follows: 
(i) the training set is used exclusively for model fitting (including model calibration and, where applicable, internal cross-validation), 
(ii) the uncertainty calibration set is used to apply the fitted model and to compute residual quantiles for uncertainty calibration, and 
(iii) the testing set is used solely to evaluate the empirical accuracy of the resulting prediction intervals \cite{gibbs2021adaptive}.

We can define the sets for model training with the sample size $n_{\mathrm{train}} = 0.4 n$, for calibration with the sample size $n_{\mathrm{cal}} = 0.4 n$ and for testing $n_{\mathrm{test}} = 0.2n$. The resulting input data matrices are denoted as $\mathbf{X}^{\mathrm{tr}}$ for training, $\mathbf{X}^{\mathrm{cal}}$ for uncertainty calibration, and $\mathbf{X}^{\mathrm{test}}$ for testing. In a simmilar manner, $\mathbf{y}^{\mathrm{tr}}$ would be the response variable in training, $\mathbf{y}^{\mathrm{cal}}$ in uncertainty calibration and $\mathbf{y}^{\mathrm{test}}$ in testing.


When appropriate, the predictor data are centered and scaled prior to model fitting. 
Centering ensures that the latent variables are defined with respect to the data mean, 
while scaling may be beneficial when predictor variables have heterogeneous magnitudes, 
so that no single variable dominates the latent-variable decomposition \cite{kettaneh2005pca}. 
Importantly, scaling is not a requirement of the proposed uncertainty calibration method and 
can be omitted in applications (e.g., certain spectroscopic datasets) where raw variable 
magnitudes carry physical meaning. The resulting preprocessed matrices are denoted 
$\widetilde{\mathbf{X}}^{\mathrm{tr}}$, $\widetilde{\mathbf{X}}^{\mathrm{cal}}$, and 
$\widetilde{\mathbf{X}}^{\mathrm{test}}$.

\subsection{Uncertainty calibration for latent-variable regression models}

In PCR, the first step is the decomposition of the data matrix into $H$ principal components (PCs):\begin{equation}
    \widetilde{\mathbf{X}} = \mathbf{T}_{\mathrm{PCA}} \mathbf{P}^\intercal_{\mathrm{PCA}} + \mathbf{E}_{\mathrm{PCA}},
    \label{eq:PCA}
\end{equation}where $\mathbf{T}_{\mathrm{PCA}}$ is the score matrix, $\mathbf{P}_{\mathrm{PCA}}$ is the loadings matrix and $\mathbf{E}_{\mathrm{PCA}}$ is the residual matrix. The loadings ($\mathbf{P}_{\mathrm{PCA}}$) are coefficients that project the original variables onto the latent axis. The data point projections are the scores ($\mathbf{T}_{\mathrm{PCA}}$). If the PCA is computed by singular value decomposition (SVD), where $ \mathbf{X} = \mathbf{U}_{\mathrm{SVD}} \mathbf{\Sigma} \mathbf{V}^\intercal_{\mathrm{SVD}}$, the loadings ($\mathbf{P}_{\mathrm{PCA}}$) are equivalent the right-hand singular vector ($\mathbf{V}_{\mathrm{SVD}}$), whereas the scores matrix ($\mathbf{T}_{\mathrm{PCA}}$) is represented by the left-hand singular vector multiplied by the singular values ($\mathbf{U}_{\mathrm{SVD}} \mathbf{\Sigma}$).

PCA is applied to the standardized training matrix, for a number of $H$ PCs. The loadings obtained are utilised to project the uncertainty calibration and test sets, as follows
\begin{equation}
    \mathbf{T}^{\mathrm{cal}}_{\mathrm{PCA}} = \widetilde{\mathbf{X}}^{\mathrm{cal}} \mathbf{P}^{\mathrm{tr}} _{\mathrm{PCA}}\text{,}
\end{equation}
\begin{equation}
    \mathbf{T}^{\mathrm{test}}_{\mathrm{PCA}} = \widetilde{\mathbf{X}}^{\mathrm{test}} \mathbf{P}^{\mathrm{tr}}_{\mathrm{PCA}} \text{.}
\end{equation}

The PC scores are then regressed against the response variable using ordinary least squares:
\begin{equation}
    \mathbf{b}_{\mathrm{PCR}} =
    \left(
    \mathbf{T}_{\mathrm{PCA}}^{\mathrm{tr}\,\intercal}
    \mathbf{T}_{\mathrm{PCA}}^{\mathrm{tr}}
    \right)^{-1}
    \mathbf{T}_{\mathrm{PCA}}^{\mathrm{tr}\,\intercal}
    \mathbf{y}^{\mathrm{tr}} .
\end{equation}

where $\mathbf{b}_{\mathrm{PCR}}$ denotes the regression coefficients in the latent-variable space.
Since the retained principal components are orthogonal by construction, the matrix 
$\mathbf{T}^{\mathrm{tr}\intercal}_{\mathrm{PCA}} \mathbf{T}^{\mathrm{tr}}_{\mathrm{PCA}}$ 
is diagonal and invertible.
 
While the uncertainty calibration procedure is demonstrated using PCR for clarity, the same steps apply unchanged to other latent-variable regression models, including PLS, K-PCR, and K-PLS, once the latent score matrix $\mathbf{T}$ has been obtained.

The LV projections of the uncertainty calibration set \textemdash here, the individual columns in $\mathbf{T}^{\mathrm{cal}}_{\mathrm{PCA}}$  \textemdash are divided into \textit{k} LV intervals. For each score vector $\mathbf{t}^{\mathrm{cal}}_h$, let $min_h$ and $max_h$ be the minimum and maximum values of $\mathbf{t}^{\mathrm{cal}}_h$. We define \textit{k} LV intervals \\ $
    \left[ min_h + \frac{i-1}{k} (max_h - min_h), min_h + \frac{i}{k}(max_h - min_h) \right], \text{ for } i = {1, 2, .., k}$. The discretized intervals of the latent variables used for residual quantile estimation are referred to as \emph{LV intervals}, 
so as not to be confused with the resulting prediction intervals, which are denoted as \emph{PI}.

For each LV interval \textit{i} of $\mathbf{t}^{\mathrm{cal}}_{h}$, we collect the indices of the sample $j$ of $\mathbf{y}^{\mathrm{cal}}$ in $I_{i,h}$, corresponding to the LV interval \begin{equation} \label{eq:quantile}
    I_{i, h} = \{ j \mid min_h + \frac{i-1}{k}(max_h - min_h) \leq \mathbf{t}_{h,j}^{\mathrm{cal}} < min_h + \frac{i}{k} (max_h - min_h)  \}.
\end{equation} 

The absolute residuals ($\mathbf{r}$) for the uncertainty calibration set are computed
\begin{equation}
    \textbf{r}^{\mathrm{cal}} = | \mathbf{y}^{\mathrm{cal}} - \mathbf{T}^{\mathrm{cal}}_{\mathrm{PCA}} \mathbf{b}_{\mathrm{PCR}}| \text{.}
\end{equation}

Then, we calculate the quantile of the prediction residuals of these samples according to our desired confidence limit. For a 0.95 confidence limit, the quantile is
\begin{equation}
    q_{i, h} = \phi^{0.95}\left(\{ \mathbf{r}^{\mathrm{cal}}[j] \mid j \in I_{i, h} \}\right),
\end{equation}
where $\phi^{0.95}(\cdot)$ denotes the empirical 0.95 quantile operator applied to the calibration residuals. 
This results in an absolute residual quantile value for each LV interval ($i$) in each LV ($h$).

To make a PI that accounts for all variables and respects the dominance of the variational profiles, a weight is calculated for each of the LVs. For each principal component \textit{h}, the explained variance in the input data is given by the eigenvalues $\lambda$, thus the LV weights ($w$) can be computed as:
\begin{equation}
    w_{h} = \frac{\lambda_h}{\sum_{i=1}^H \lambda_i} \text{.}
\end{equation}

When bringing new data into the model, the first step is to project the data onto the LVs
\begin{equation}
    \mathbf{T}^{\mathrm{test}}_{\mathrm{PCA}} = \widetilde{\mathbf{X}}^{\mathrm{test}}\mathbf{P}_{\mathrm{PCA}}^{\mathrm{tr}} \text{,}
\end{equation}
and utilize the model to make estimations
\begin{equation}
    \hat{\mathbf{y}}^{\mathrm{test}} = \mathbf{T}_{\mathrm{PCA}}^{\mathrm{test}} \mathbf{b}_{\mathrm{PCR}} \text{.}
\end{equation}

For each LV \textit{h} and each test sample \textit{j}, we determine membership to the calibration LV intervals ($I_{i,h}$), and retrieve the weighted quantile for that LV interval from calibration ($q_{j}$)
\begin{equation}
    q_{j} = \sum^{H}_{h=1} w_h q_{I_{i,h}}
    \label{eq:weight}
\end{equation}
where $I_{i,h}$ is the LV interval index where the test sample \textit{j} maps in the \textit{h}-th LV. 

If a test sample projects outside the range spanned by the calibration scores in a given latent variable, i.e.,
$t_{h,j}^{\mathrm{test}} < \min(\mathbf{t}^{\mathrm{cal}}_h)$ or
$t_{h,j}^{\mathrm{test}} > \max(\mathbf{t}^{\mathrm{cal}}_h)$,
the sample is assigned to the nearest boundary interval ($i=1$ or $i=k$, respectively).
In other words, the outermost LV intervals are treated as open-ended intervals.

\begin{figure}[H]
    \centering
\includegraphics[width=0.9\linewidth]{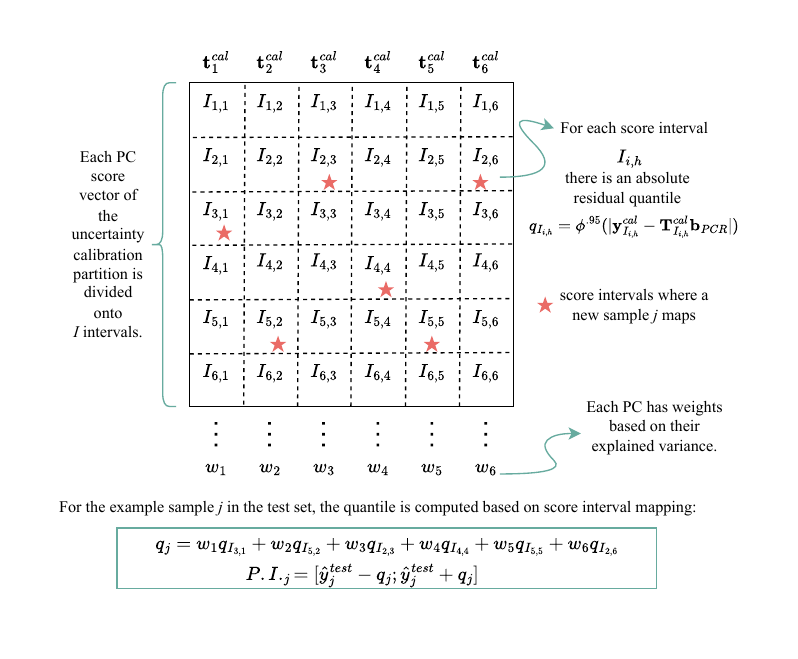}
\caption{The diagram represents a visualization of the PI calculation for a new test sample $j$, where a model with 6 PCs and 6 score LV intervals is chosen.}
\label{fig:workflowTestConfidence}
\end{figure}

The process is exemplified in Figure \ref{fig:workflowTestConfidence}. For a given test sample $j$, this equation aggregates information from all $H$ latent variables to produce a single residual quantile that defines the width of the prediction interval. Specifically, the test sample is first projected onto each latent variable $h = 1,\ldots,H$. Based on its score value in the $h$-th latent variable, the sample is assigned to one of the pre-defined calibration intervals $I_{i,h}$. From the calibration step, each such interval is associated with an empirical residual quantile $q_{i,h}$, estimated from calibration samples whose latent-variable projections fall within the same interval.

The quantity $q_{I_{i,h}}$ therefore represents the expected magnitude of the prediction residual for samples that occupy a similar region of the latent space as the test sample in latent variable $h$. To combine the uncertainty contributions across all latent variables, these interval-specific quantiles are aggregated using a weighted sum, where the weights $w_h$ are proportional to the explained variance of each latent variable.

As a result, latent variables that capture a larger fraction of the systematic variation in the predictor data contribute more strongly to the final prediction interval, while latent variables with smaller explained variance have a reduced influence. The resulting value $q_j$ can be interpreted as an estimate of the expected absolute prediction error for test sample $j$, informed by its position in the latent space across all retained latent variables.

The use of a weighted sum in Eq.~\refeq{eq:weight}  is motivated by the additive nature of explained variance in latent-variable models. In PCA- and PLS-based decompositions, the total variance in the predictor space is partitioned across orthogonal latent variables, and the explained variances of these latent variables sum to the total explained variance. In the proposed method, this property is used to define weights that reflect the relative importance of each latent variable when aggregating uncertainty information. The explained variance is used as a principled weighting mechanism to combine latent-variable–specific residual quantiles into a single prediction-interval half-width. This results in an uncertainty estimate that is dominated by systematic latent structures while limiting the influence of noise-dominated components.

The use of a weighted aggregation is also consistent with classical concepts of uncertainty propagation in analytical chemistry. According to IUPAC \cite{iupac_orangebook_2014} guidelines, when an output quantity depends on multiple intermediate quantities, the overall uncertainty can be approximated by combining the individual contributions according to their relative sensitivities. In LV models, the orthogonality of the LVs and the partitioning of systematic variance across components provide an analogous structure: each LV represents a distinct source of systematic variation, and its explained variance reflects its relative contribution to model behavior. 

Intuitively, this procedure can be viewed as a consensus estimate of uncertainty across multiple latent directions. Each latent variable provides an independent, localized estimate of prediction uncertainty based on past calibration samples with similar latent projections. The weighted aggregation ensures that uncertainty is driven primarily by systematic latent structures rather than by noise-dominated components.

The last step is to construct PI for each sample 
\begin{equation} \label{eq:interv}
    PI_{j} = \left[ \hat{y}^{\mathrm{test}}_{j} - q_{j};  \hat{y}^{\mathrm{test}}_{j} + q_{j}\right] \textbf{.}
\end{equation}

A visualisation of the workflow for PCR is presented in \textbf{Appendix A}. For PLS, K-PCR, and K-PLS, the uncertainty calibration is performed identically, conditioned on the corresponding latent-variable score matrices. Since these regression models are well established, their training and latent-space construction are not repeated here. Interested readers are referred to standard references \cite{wold2001pls, rosipal2001kernel, hoegaerts2005subset} and to \textbf{Appendix~B} for implementation details relevant to kernel-based models.

\section{Case Studies}

\begin{figure}[H]
    \centering
    \includegraphics[width=0.8\linewidth]{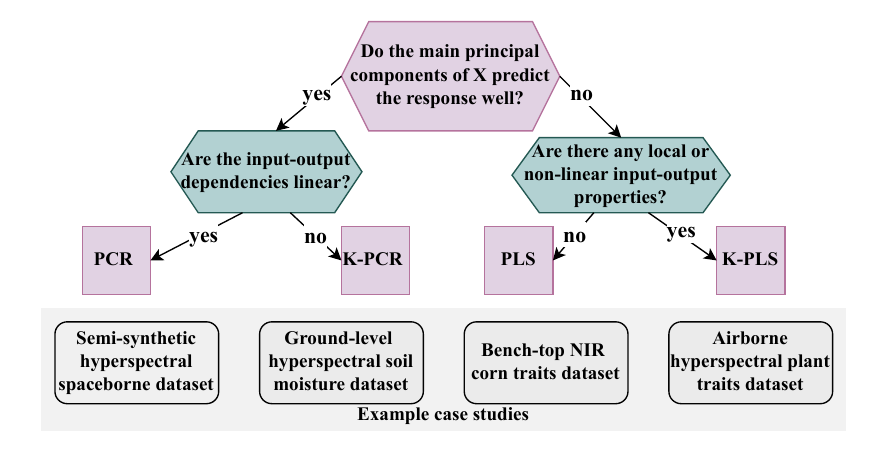}
    \caption{The rationale for choosing the models, and the example case studies presented for each scenario.}
    \label{fig:modelDecision}
\end{figure}

Each multivariate statistical tool presented, along with their kernelized versions, is suitable for specific scenarios, as illustrated in Figure~\ref{fig:modelDecision}. 
The principal difference between PCR and PLS models lies in how they relate the explanatory variables to the response variable. 
PCR may be misleading when the directions of highest variance in the explanatory variables (\textbf{X}) do not correspond to directions that explain the response (\textbf{y}) well (i.e., low $\mathrm{R}^2_{\text{Y|X}}$). 
PLS addresses this by rotating the latent variables toward the directions of maximum covariance between \textbf{X} and \textbf{y}, such that the resulting latent variables $\mathbf{T}_{\mathrm{PLS}}$ are directly optimized to predict \textbf{y}. 
In other words, when the main principal components of \textbf{X} are not predictive of \textbf{y}, PLS is more advantageous.

The bilinear multivariate methods are optimal when the relationships between explanatory and response variables exhibit linear dependencies. Only mildly non-linear phenomena can be modeled with these linear methods. If the dependencies are strongly non-linear or measurement interference causes non-linearity, the kernelized versions of PLS and PCR should be applied.

Each case study, in addition to illustrating the proposed uncertainty calibration procedure, is designed to examine the effect of a specific modeling or calibration choice (e.g., number of latent variables or discretization resolution) on the resulting prediction intervals, as summarized in Table~\ref{tab:placeholder}.

\begin{table}[H]
    \centering
    \caption{The particular scenarios tested with each case study.}
    \begin{tabular}{c|c|c}
       \textbf{Case study} & \textbf{Model type} & \textbf{Testing} \\ \hline
        Case 1 & PCR & The correctness of the estimated uncertainty; \\
        Case 2 & K-PCR & The influence of the number of LVs and LV intervals;  \\ 
        Case 3 & PLS & The influence of the number of LV intervals; \\ 
        Case 4 & K-PLS & The accuracy of the PI . \\ 
    \end{tabular}
    \label{tab:placeholder}
\end{table}

\subsection{Case 1: Semi-synthetic hyperspectral spaceborne data}

To evaluate the ability of the proposed method to localize and calibrate predictive uncertainty, we construct a semi-synthetic regression problem in which the true structure of uncertainty is known by design. This allows a direct and interpretable assessment of whether the method assigns wider prediction intervals to regions of genuinely higher uncertainty.

\begin{figure}[H]
    \begin{subfigure}[b]{0.62\linewidth}
        \centering
        \includegraphics[width=0.8\linewidth]{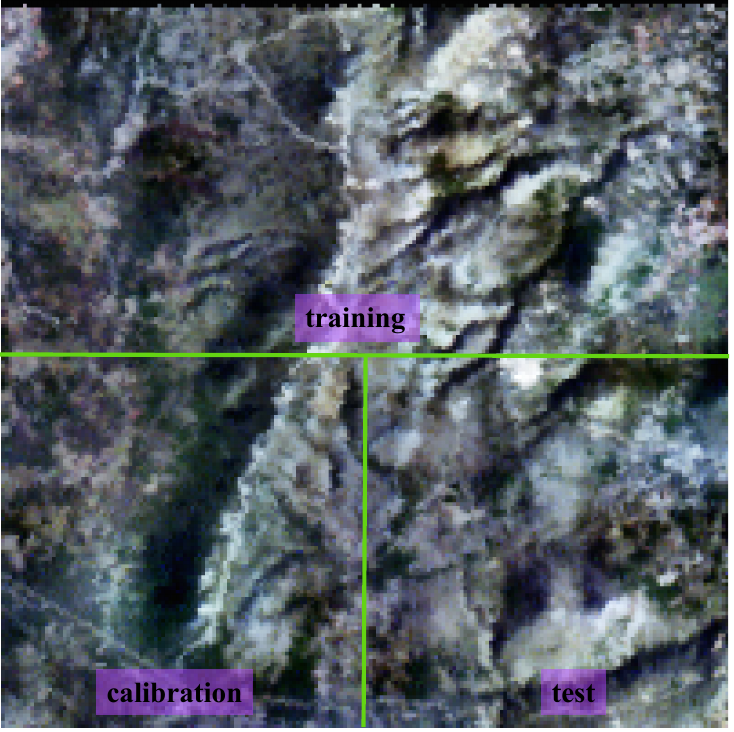}
        \caption{}
        \label{fig:hyperCube}
    \end{subfigure}
    \hfill
    \centering
        \begin{subfigure}[b]{0.62\linewidth}
        \centering
        \includegraphics[width=0.9\linewidth]{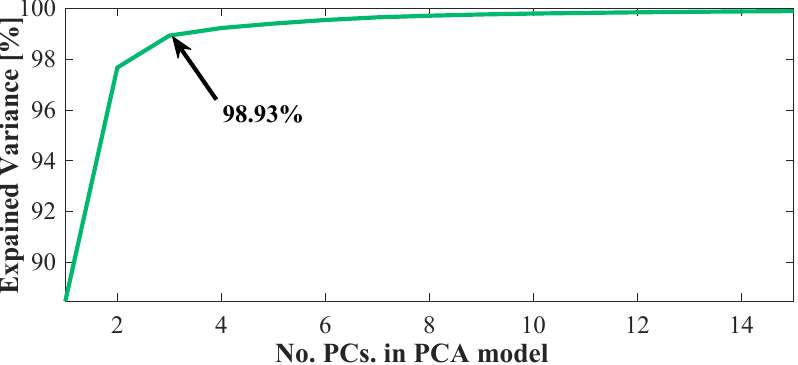}
        \caption{}
        \label{fig:varianceExplained}
    \end{subfigure}
    \caption{(a) The PRISMA image utilised in the semi-synthetic case (Case 1) along with (b) the explained variance of PCs for the training set.}
    \label{fig:case1cube}
\end{figure}

The input data consist of a real hyperspectral satellite image acquired by the PRISMA mission (Fig.~\ref{fig:hyperCube}). The image is atmospherically corrected (L2D processing level) and contains 66 bands in the visible and near-infrared range. The scene is spatially divided into three disjoint regions used as training (50\%), uncertainty calibration (25\%), and test sets (25\%), respectively. PCA is applied to the training set, and the first three principal components explain approximately 99\% of the total variance of the training image (Fig.~\ref{fig:varianceExplained}).

\begin{figure}[H]    
    \centering
    \begin{subfigure}[b]{0.49\linewidth}
        \centering
        \includegraphics[width=\linewidth]{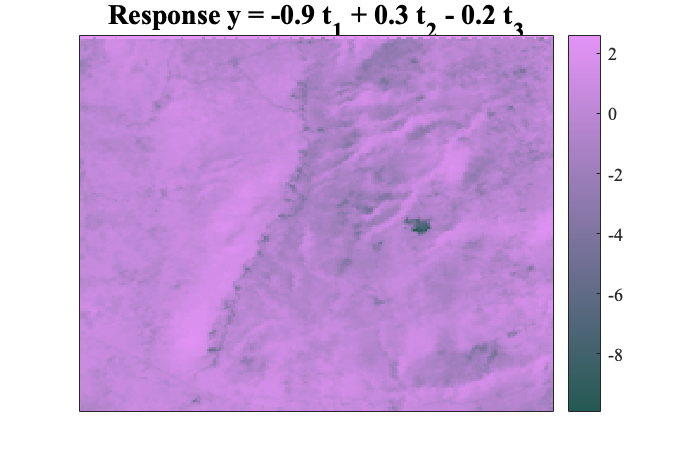}
        \caption{$\mathbf{y}$}
        \label{fig:responseCase4}
    \end{subfigure}
    \begin{subfigure}[b]{0.49\linewidth}
        \centering
        \includegraphics[width=\linewidth]{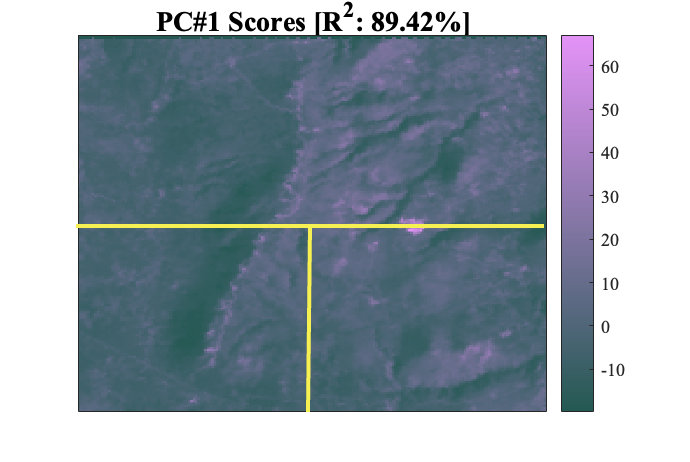}
        \caption{$\mathbf{t}_1$}
        \label{fig:PC1t}
    \end{subfigure}
    \begin{subfigure}[b]{0.49\linewidth}
        \centering
        \includegraphics[width=\linewidth]{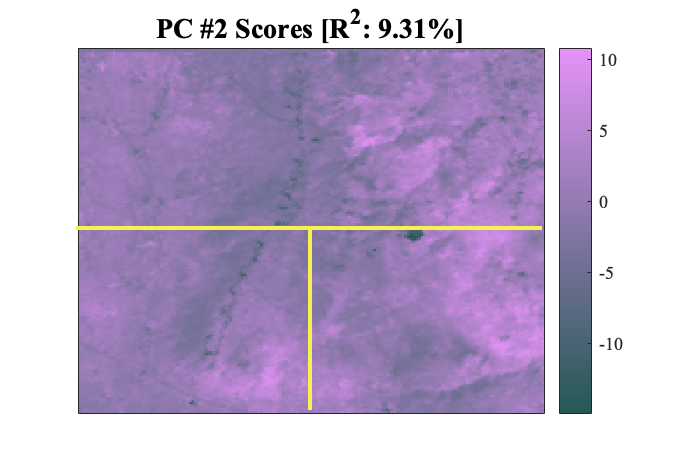}
        \caption{$\mathbf{t}_2$}
        \label{fig:PC2t}
    \end{subfigure}
    \begin{subfigure}[b]{0.49\linewidth}
        \centering
        \includegraphics[width=\linewidth]{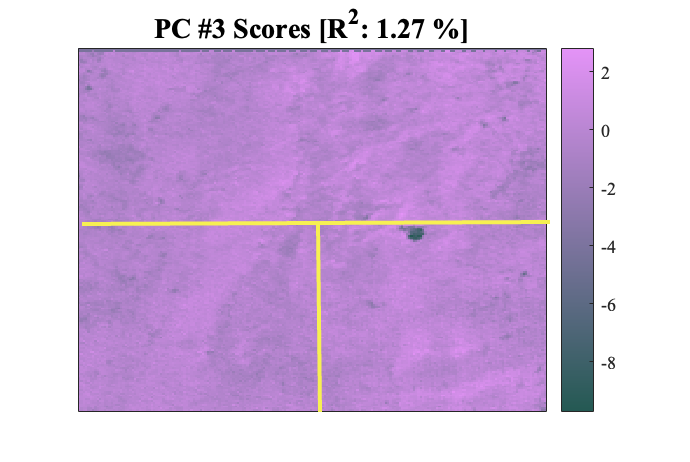}
        \caption{$\mathbf{t}_3$}
        \label{fig:PC3t}
    \end{subfigure}
    \caption{(a) The synthethic response variable, prior to noise addition, along with the (b)-(d) principal component score (\textbf{T}) maps.}
    \label{fig:descriptionCase1}
\end{figure}

A synthetic response variable is then constructed as a linear combination of the first three principal component score maps,
\begin{equation}
\mathbf{y} = -0.9,\mathbf{t}_1 + 0.3,\mathbf{t}_2 - 0.2,\mathbf{t}_3 ,
\end{equation}
where $\mathbf{t}_1$, $\mathbf{t}_2$, and $\mathbf{t}_3$ denote the corresponding PC scores (Fig.~\ref{fig:descriptionCase1}). This response is noiseless by construction and therefore perfectly predictable from the retained latent variables.

To introduce controlled, input-dependent uncertainty, heteroscedastic noise is added to the response variable. The noise magnitude is defined as a function of the first principal component score: samples with negative PC1 values are assigned low noise levels, while samples with large positive PC1 values receive substantially higher noise. This design creates spatially structured uncertainty that depends explicitly on the latent-variable representation of the input data. The resulting semi-synthetic dataset therefore combines realistic hyperspectral predictors with a known and interpretable uncertainty pattern.

A successful uncertainty calibration should assign wider prediction intervals to regions where the model exhibits larger prediction errors. Importantly, this relationship is not enforced by construction: the residual quantiles are computed exclusively from calibration data and depend only on latent-variable projections, not on test residuals.

The objective of this case study is to assess whether the proposed LV-based uncertainty calibration method can (i) correctly identify regions of increased predictive uncertainty using only input information and (ii) translate this information into locally adaptive PIs for unseen test samples. In addition, PIs obtained with the proposed method are compared to global split conformal and bootstrap-based baselines to highlight the benefits of latent-space localization.

\subsection{Case 2: Soil moisture regression with ground-level hyperspectral data}

The soil moisture dataset acquired by Riese and Kellar, 2018 \cite{riese2018hyperspectral} originates from a five-day campaign where undisturbed soil samples were measured both with a hyperspectral camera and ground-level moisture content. The observations have measured reflectace in 125 spectral bands between 450 and 950 nm, and a spatial resolution of 4 nm. 
The authors have previously demonstrated in Duma \textit{et al.}, 2024 the appropriateness to utilize Kernel-PCR for the task, as the most important PCs of the input spectral data are correlated with soil moisture $\left[ \% \right]$.

\subsection{Case 3: Corn property regression models with bench-top NIR spectroscopy}

\begin{figure}[H]
    \centering
    \begin{subfigure}[b]{0.49\linewidth}
        \centering
        \includegraphics[width=0.99\linewidth]{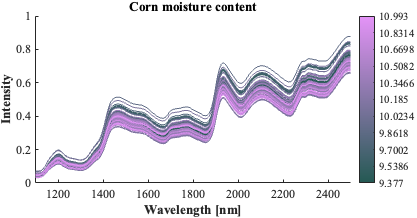}
        \caption{}
        \label{fig:moisture}
    \end{subfigure}
    \hfill
    \begin{subfigure}[b]{0.49\linewidth}
        \centering
        \includegraphics[width=0.99\linewidth]{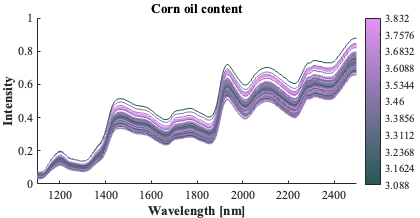}
        \caption{}
        \label{fig:oil}
    \end{subfigure}
    \hfill
    \begin{subfigure}[b]{0.49\linewidth}
        \centering
        \includegraphics[width=0.99\linewidth]{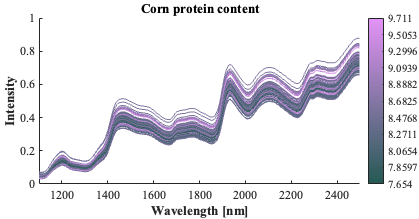}
        \caption{}
        \label{fig:protein}
    \end{subfigure}
    \hfill
    \begin{subfigure}[b]{0.49\linewidth}
        \centering
        \includegraphics[width=0.99\linewidth]{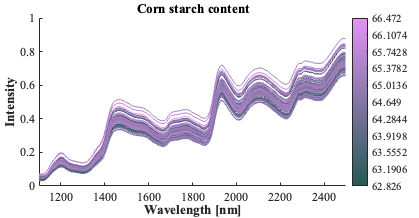}
        \caption{}
        \label{fig:starch}
    \end{subfigure}
    \caption{Corn dataset measured by the m5 spectrometer [data source: http://www.eigenvector.com/data/Corn
]}
    \label{fig:case2data}
\end{figure}

The benchmark NIR dataset contains corn spectra measured with different NIR instruments. The corresponding quantitative properties of the corn to be regressed are contents of moisture, oil, protein and starch. The utilised dataset is visualized in Fig. \ref{fig:case2data}. The aim of the case study is to illustrate the proposed uncertainty calibration procedure on PLS regression, as the Corn dataset has been a benchmark for testing various PLS variants.

\subsection{Case 4: Plant trait regression models with airborne hyperspectral data}

The hyperspectral K-PLS regression model for plant traits estimation is calibrated with airborne hyperspectral reflectance data collected during the SHIFT campaign \cite{chadwick2025unlocking} using an AVIRIS-NG instrument \cite{brodrick2023shift} and ground-level measurements for Leaf Mass Area (LMA) \cite{zheng2024shift}. The reflectance (L2A) is in visible to shortwave infrared (VSWIR) range, with wavelengths between 380-2150 nm and a spectral resolution of 5 nm. 

\section{Results and discussion}

\subsection{Case 1 results and discussion on uncertainty calibration accuracy}

\begin{figure}[H]
    \centering
    \begin{subfigure}[b]{0.49\linewidth}
        \centering
        \includegraphics[width=\linewidth]{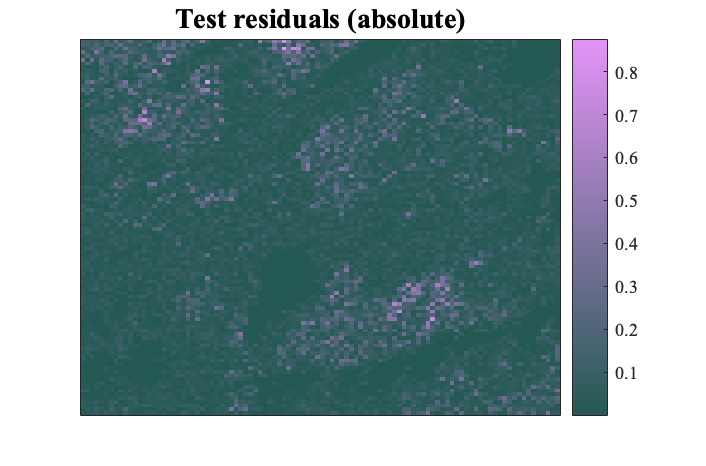}
        \caption{}
        \label{fig:residualsCase1}
    \end{subfigure}
    \hfill
    \begin{subfigure}[b]{0.49\linewidth}
        \centering
        \includegraphics[width=\linewidth]{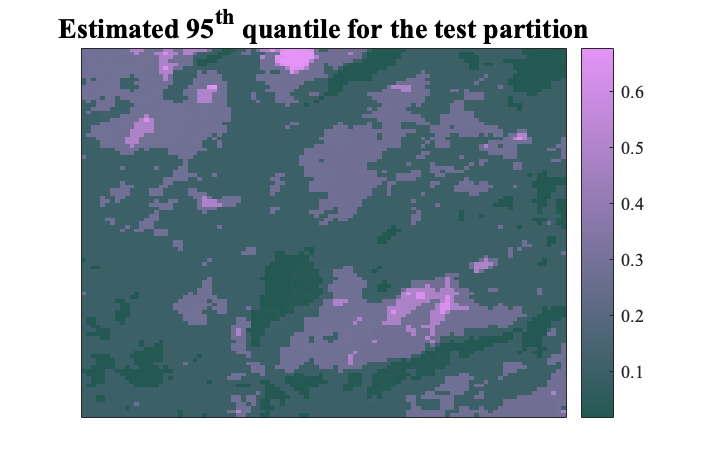}
        \caption{}
        \label{fig:quantileCase1}
    \end{subfigure}
    \caption{(a) The residuals of the test set, along with (b) the estimated residual $95^{th}$ quantile. Pixel-level relationship between absolute test residuals $|e_j|$ and the estimated 95\% residual quantiles $q_j$ for the test set in Case~1. The positive Pearson ($r = 0.65$) and Spearman ($\rho = 0.55$) correlations confirm that higher prediction errors are associated with wider prediction intervals, supporting the ability of the proposed method to localize uncertainty in latent space.}
    \label{fig:syntheticModel}
\end{figure}

Case~1 evaluates whether the proposed LV-localized uncertainty calibration can recover a known, LV–dependent heteroscedastic uncertainty pattern in a controlled semi-synthetic setting. Figure~\ref{fig:syntheticModel} compares the absolute prediction residuals on the test set (Fig.~\ref{fig:residualsCase1}) with the corresponding LV-localized estimated 95\% residual quantiles $q_j$ (Fig.~\ref{fig:quantileCase1}), computed solely from calibration residuals and the latent projections of the inputs.

To quantitatively support the visual interpretation, we computed the pixel-level association between the absolute test residuals $|e_j|$ and the estimated 95\% residual quantiles $q_j$. A clear positive relationship was observed (Pearson correlation coefficient $r = 0.65$, Spearman rank correlation $\rho = 0.55$, both $p \ll 0.001$), indicating that samples with larger prediction errors are systematically assigned wider prediction intervals. This behavior is expected in the present semi-synthetic experiment, where the noise magnitude was constructed to vary as a function of the LVs. The observed correlation therefore provides quantitative evidence that the proposed uncertainty calibration procedure successfully recovers spatially varying uncertainty patterns encoded in the latent space.

To contextualize the proposed LV-localized prediction intervals, Table~\ref{tab:case1_baselines} compares their performance against three commonly used baseline approaches: global split conformal prediction, residual bootstrap prediction intervals, and a naïve (pairs) bootstrap procedure in which training samples are resampled and the full PCR model is refitted at each replicate.

The proposed LV-localized method achieves empirical coverage of 95.23\%, closely matching the nominal 95\% target, while maintaining moderate prediction interval widths (mean width 0.5628, Table~\ref{tab:case1_baselines}). In contrast, both the global split conformal and residual bootstrap methods substantially under-cover, achieving only 90.16\% and 90.02\% coverage, respectively, despite producing similarly narrow intervals (mean widths $\approx 0.504$).
Methods based on a single global residual distribution cannot adapt interval widths to regions of higher uncertainty, resulting in overly narrow prediction intervals in those regions.

The naïve (pairs) bootstrap performs particularly poorly, yielding essentially zero empirical coverage (0.02\%) together with vanishingly small prediction intervals (mean width $4.4\times10^{-4}$). This failure arises because resampling training pairs and recomputing PCA leads to strong shrinkage of the effective noise in this controlled setting: the dominant latent structure is repeatedly re-estimated from highly similar resampled datasets, while the heteroscedastic noise is not properly represented in the bootstrap refits. As a result, the resulting prediction distribution severely underestimates predictive uncertainty. In addition to its lack of validity in this scenario, the naïve bootstrap is computationally prohibitive, requiring orders of magnitude more computation time (78.8~s) than the other methods.

Overall, this comparison highlights the practical advantage of the proposed LV-localized approach. Compared to naïve bootstrap methods, it avoids repeated model refitting and provides well-calibrated prediction intervals at a fraction of the computational cost (0.023~s vs. 78.8~s). Compared to global residual-based methods, it adapts uncertainty to the LV structure of the data, restoring near-nominal coverage under input-dependent and heteroscedastic uncertainty (95.23\% vs. $\approx$90\%). These results demonstrate that the proposed method occupies a favorable trade-off between calibration accuracy, adaptivity, and computational efficiency.

\begin{table}[H]
\centering
\caption{Case~1: Empirical test coverage (nominal 95\%), prediction interval (PI) width statistics, and computational time for the proposed method and baseline uncertainty quantification approaches. Exact mathematical definitions and hyperparameter settings are provided in \textbf{Appendix~C}.}
\label{tab:case1_baselines}
\begin{tabular}{lrrrr}
\toprule
Method & Coverage [\%] & Mean PI width & Median PI width & Time [s] \\
\midrule
LV-localized PI (proposed) & 95.2300 & 0.5628 & 0.4811 & 0.0232 \\
Global split conformal PI  & 90.1600 & 0.5046 & 0.5046 & 0.0016 \\
Residual bootstrap PI      & 90.0200 & 0.5043 & 0.5035 & 0.7267 \\
Na\"ive (pairs) bootstrap PI & 0.0200 & 0.0004 & 0.0004 & 78.7680 \\
\bottomrule
\end{tabular}
\end{table}

Consistent with Table~\ref{tab:case1_baselines}, the resulting LV-localized prediction intervals achieve near-nominal empirical coverage on the test set while adapting interval width to the latent-variable–dependent error structure.

\subsection{Case 2 results and discussion on LV selection}

\begin{figure}[H]
    \centering
    \includegraphics[width=0.8\linewidth]{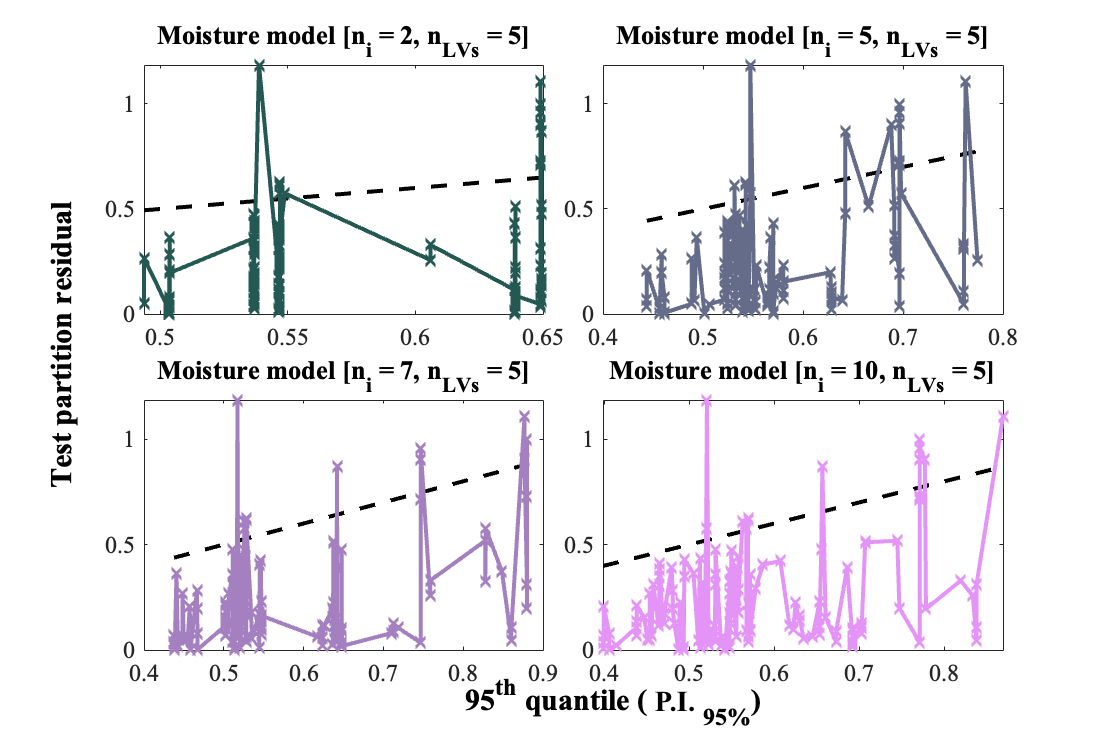}
    \caption{The test set residual magnitude and the estimated $95^{th}$ quantile value for the observation, for the 5 LV K-PCR model utilising 2, 5, 7 and 10 LV intervals.}
    \label{fig:soilSamplesOver}
\end{figure}

Case~2 examines how the uncertainty calibration depends on discretization resolution ($k$) and the number of retained latent variables ($H$). Figure~\ref{fig:soilSamplesOver} visualizes test-set absolute residuals against the corresponding estimated 95\% residual quantiles $q_j$. In this diagnostic, points above the line $|e_j| = q_j$ represent samples whose residuals exceed the nominal 95\% prediction-interval half-width; under correct calibration, approximately 5\% of test samples are expected to lie above this line.

When the number of LV intervals is too small (e.g., $k=2$), calibration becomes overly global and uncertainty is misrepresented: many high-error samples exceed the nominal threshold, while some low-error samples receive overly large quantiles. Increasing $k$ improves localization and brings exceedances closer to the nominal 5\% level, until overly fine discretization leads to sparse intervals and unstable quantile estimates.

The estimated quantile is determined based on the LV interval to which each test sample is assigned. It is observed that if the number of latent variable intervals is too little (i.e. 2 LV intervals), (a) the amount of samples that exceed the 95\% PI exceed 5\% of the samples, and (b) samples with a low residual reside in a large quantile (overestimating the PI). However, as the number of LV intervals is increased (i.e. 10 LV intervals), one can observe that (a) the points are closer to the diagonal line, but residing under it, meaning the uncertainties are not overestimated and (b) the number of points over the 95\% PI does not exceed 5\%.

\begin{figure}[H]
    \centering
    \hfill
    \begin{subfigure}[b]{0.49\linewidth}
        \includegraphics[width=\linewidth]{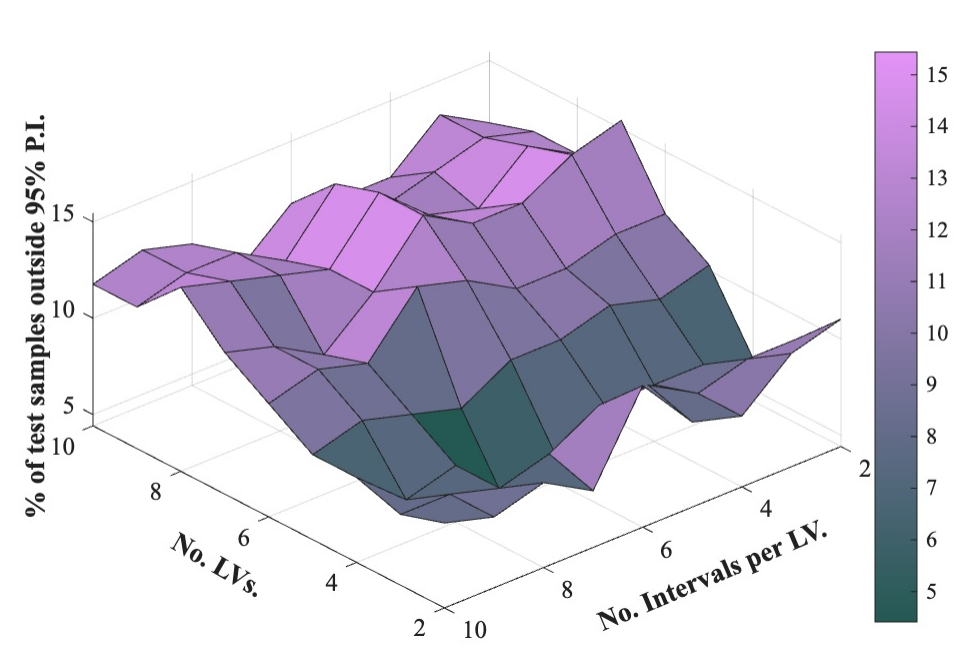}
        \caption{}
        \label{fig:soilPercentageOver}
    \end{subfigure}
    \hfill
    \begin{subfigure}[b]{0.49\linewidth}
        \includegraphics[width=\linewidth]{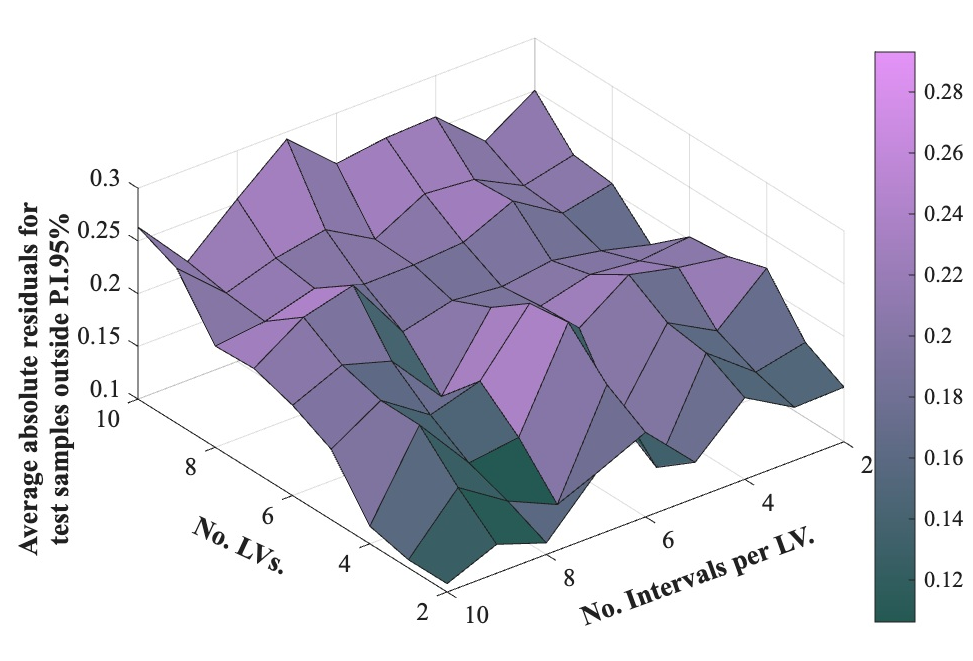}
        \caption{}
        \label{fig:soilValueOver}
    \end{subfigure}
    \caption{(a) The accuracy of the prediction confidence, measured as the percentage of absolute residual values for the test set exceeding the $95\%$ CL. 5\% of the observations are expected to exceed the PIs. (b) The average values for the absolute residuals of the samples that are exceeding the PIs. }
    \label{fig:soilModel}
\end{figure}

Fig.~\ref{fig:soilModel}, shows that for the uncertainty calibration of the K-PCR model, there is an optimal value of 4 LVs and 7 LV intervals that give a maximum of 5\% of the test samples to reside outside of the 95\% PI. The number of LVs is chosen based on cross-validation statistics, and the results along the LV axis are visualised here for further analysis. It is important to mention that the convergence of optimal uncertainty calibration in 5 LVs matches the 5-fold cross-validation results for the K-PCR model. 

In Case 2, it can be observed that the larger the number of LV intervals, the better the fit of the uncertainty calibration, until it reaches an optimum. If this optimal number of 7 LV intervals is surpassed, it can result in LV intervals within the calibration set that lack sufficient samples (overdiscretization), leading to erroneous quantile estimation and ultimately reducing the quality of uncertainty calibration. 

The number of LVs is a more significant factor in determining the accuracy of the PI. When keeping the number of LVs fixed, varying the number of LV intervals has a smaller impact on calibration compared to the reverse scenario. For instance, if we keep the number of LVs constant (to 4) and only change the number of LV intervals, the percentage of samples above the PI varies between 4.8\% and 6.2\%. However, if we maintain the number of LV intervals (at 8) and vary the number of LVs, the percentage of samples above the PI can range from 4.8\% to 13.7\%. 

This case study demonstrates that while discretization resolution influences uncertainty calibration, the number of retained latent variables plays a more dominant role, emphasizing the importance of appropriate LV selection for reliable prediction intervals.

\subsection{Case 3 results and discussion on the discretization resolution influence}

From the NIR data, four corn traits were retrieved using PLS models: moisture, oil, protein, and starch content. The predictive model was calibrated following standard practice, with the number of LVs selected exclusively based on predictive performance using 5-fold cross-validation. A combined assessment of the root mean square error during cross-validation ($RMSE_{\mathrm{CV}}$) and the coefficient of determination ($Q^{2}_{\mathrm{CV}}$) indicated that 5 LVs provide an appropriate bias–variance compromise for the moisture model, which is analyzed further in this case study.

\begin{figure}[H]
    \centering
    \hfill
    \begin{subfigure}[b]{0.48\linewidth}
        \includegraphics[width = 0.99\linewidth]{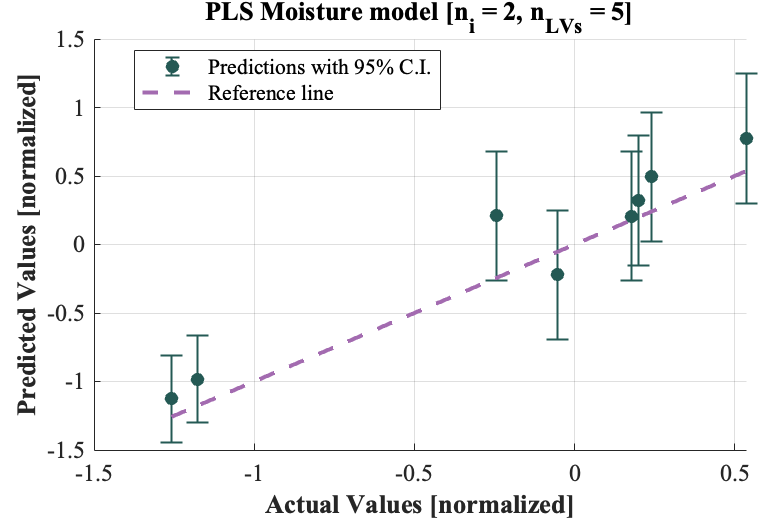}
        \caption{}
        \label{fig:2ni5LVsCorn}
    \end{subfigure}
    \hfill
    \begin{subfigure}[b]{0.48\linewidth}
        \includegraphics[width = 0.99\linewidth]{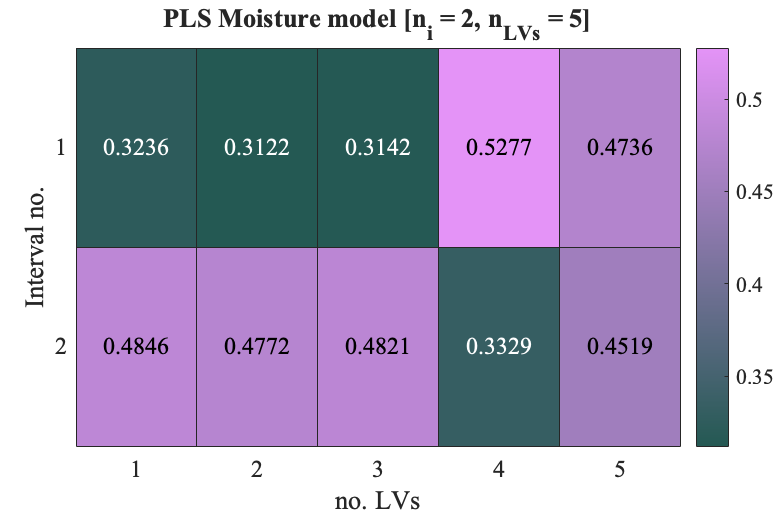}
        \caption{}
        \label{fig:2ni5LVsCornQuantile}
    \end{subfigure}
     \hfill
    \begin{subfigure}[b]{0.48\linewidth}
        \includegraphics[width = 0.99\linewidth]{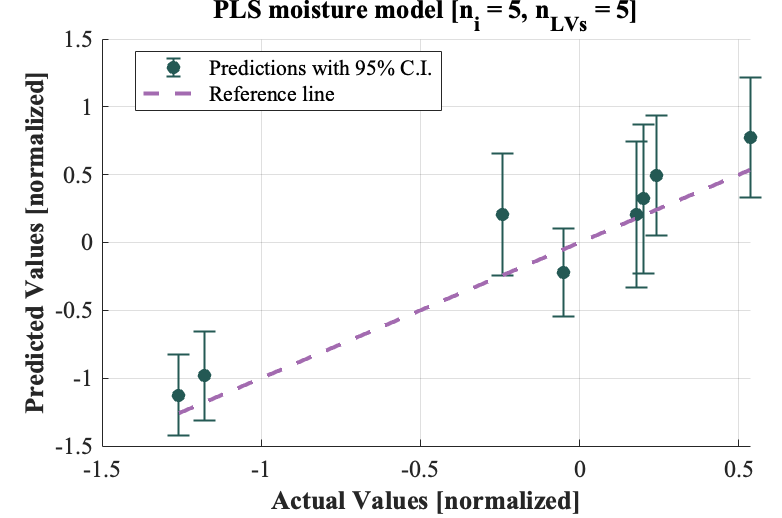}
        \caption{}
        \label{fig:5ni5LVsCorn}
    \end{subfigure}
    \hfill
    \begin{subfigure}[b]{0.48\linewidth}
        \includegraphics[width = 0.99\linewidth]{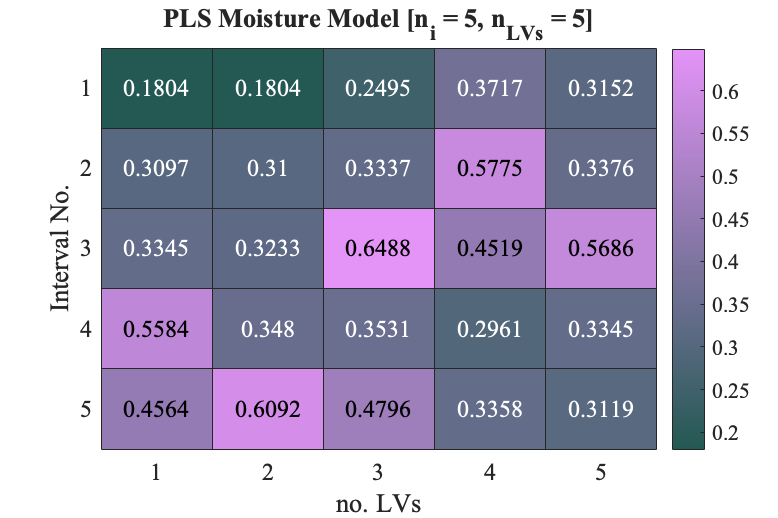}
        \caption{}
        \label{fig:5ni5LVsCornQuantile}
    \end{subfigure}
    \hfill
    \begin{subfigure}[b]{0.48\linewidth}
        \includegraphics[width = 0.99\linewidth]{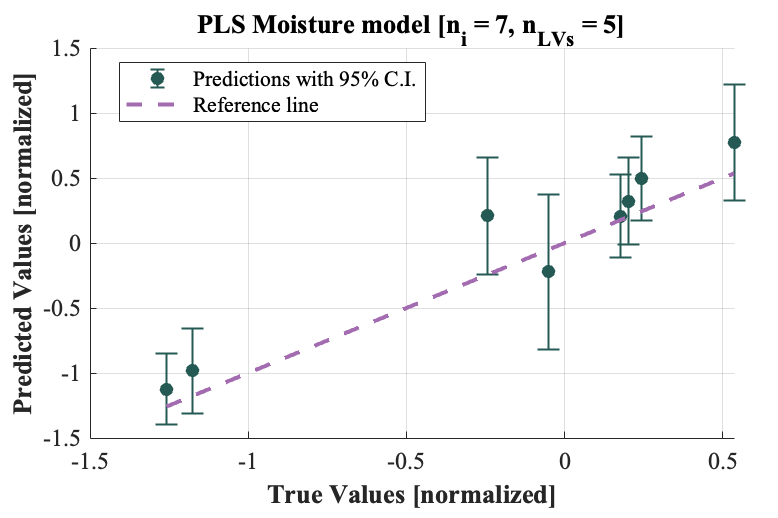}
        \caption{}
        \label{fig:7ni5LVsCorn}
    \end{subfigure}
    \hfill
    \begin{subfigure}[b]{0.48\linewidth}
        \includegraphics[width = 0.99\linewidth]{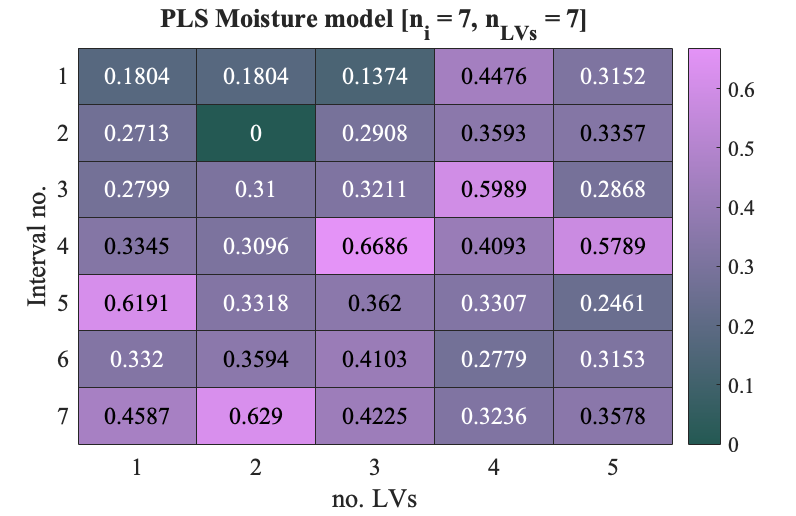}
        \caption{}
        \label{fig:7ni5LVsCornQuantile}
    \end{subfigure}
    \caption{The estimation with PIs for the number of LV intervals equal to 2, 5 and 7.}
    \label{fig:PLSResults}
\end{figure}

Figure~\ref{fig:PLSResults} evaluates the effect of the LV discretization resolution on PI construction, while keeping the predictive model fixed. Across all discretization settings, all test samples fall within the resulting PIs, indicating conservative coverage in this well-behaved laboratory dataset.

\begin{figure}[H]
    \centering
    \includegraphics[width = 0.9\linewidth]{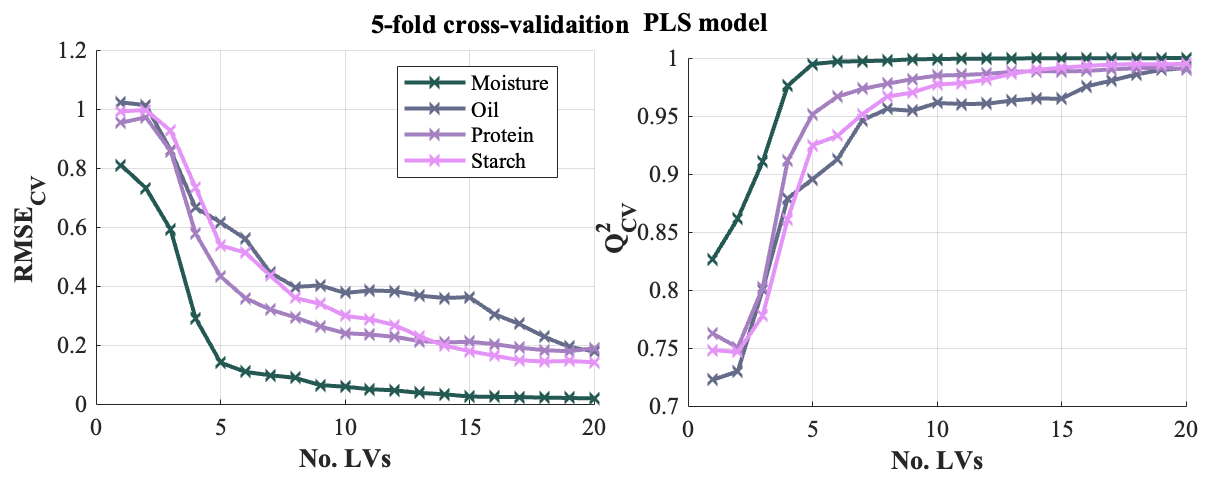}
    \caption{The PLS model performance for corn moisture estimation in the left-out test set.}
    \label{fig:PLSCornModel}
\end{figure}

When the number of LV intervals is very small, the PIs become excessively wide and show limited spatial variation (Fig.~\ref{fig:2ni5LVsCorn}). This behavior reflects the fact that coarse discretization pools calibration residuals across heterogeneous regions of the latent space, leading to overly global uncertainty estimates. Correspondingly, the residual quantiles within each LV interval exhibit little variation (Fig.~\ref{fig:2ni5LVsCornQuantile}).

As the number of LV intervals increases, the residual quantiles become more localized and exhibit greater variability across the latent space (Fig.~\ref{fig:7ni5LVsCornQuantile}). This results in sharper prediction intervals that better reflect local uncertainty without sacrificing coverage (Fig.~\ref{fig:7ni5LVsCorn}). Importantly, this refinement does not alter the underlying predictive model, but improves the representativeness of the residual quantiles used for uncertainty calibration.

These results illustrate a key practical consideration: the LV discretization should be sufficiently fine to capture heterogeneity in the residual structure, while ensuring that each interval contains enough calibration samples to yield stable quantile estimates.

These results indicate that, for well-behaved laboratory datasets, the proposed calibration method yields stable and conservative prediction intervals, with discretization primarily affecting interval sharpness rather than coverage.

\subsection{Case 4 results and discussion on prediction interval accuracy}

\begin{figure}[H]
    \centering
    \includegraphics[width=0.6\linewidth]{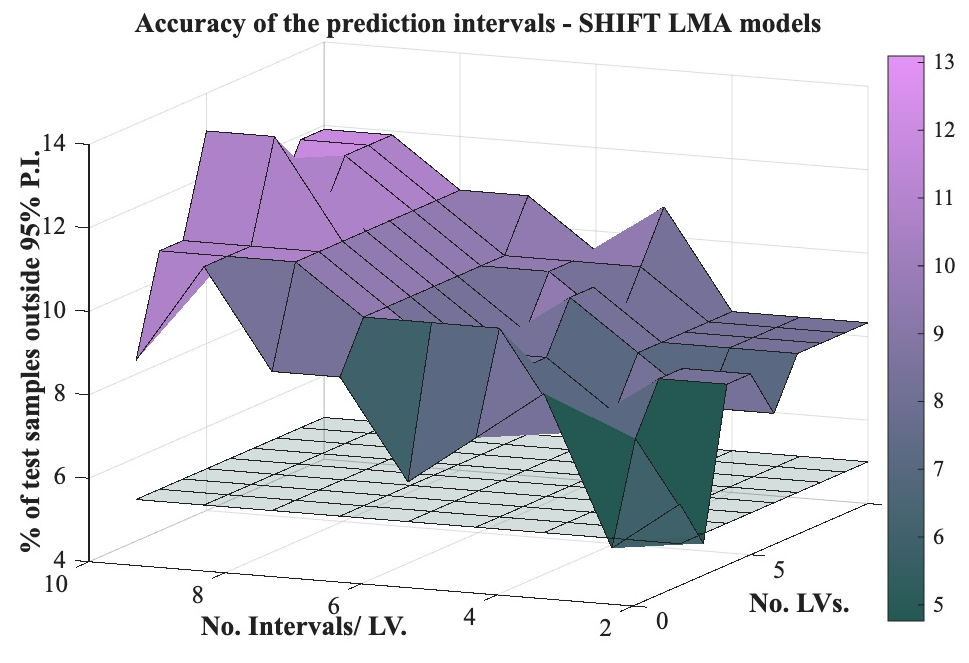}
    \caption{The accuracy of the prediction intervals for the SHIFT LMA models. The plane marks the expected 5\% of the samples to exceed the PIs.}
    \label{fig:SHIFTIntervals}
\end{figure}

Case~4 represents a challenging airborne remote-sensing setting, where substantial uncertainty arises from imperfect spectrum--trait pairing, spectral mixing at the pixel scale, and measurement noise. In this high-uncertainty regime, Fig.~\ref{fig:SHIFTIntervals} indicates that the best-calibrated prediction intervals are obtained with a small number of retained LVs and a coarse LV discretization.

From Fig.~\ref{fig:SHIFTIntervals}, it can be seen that the optimal uncertainty calibration yields a low number of LVs (2 LVs) and a small number of discrete LV intervals (2-4 LV intervalss). The low-LV model is justified by the fact that the PLS model is in the reproducing kernel Hilbert space 
$\mathscr{H}$. When data is initially mapped into a higher-dimensional space where the dependencies between spectra and traits are linear, the number of LVs required for model training decreases. This is because later LVs are not needed to explain potential deviations from the linear dependency. The optimal number of LVs by 5-fold cross-validation for the LMA model is 4, whereas the optimal number of LVs for uncertainty calibration is 2. 

Moreover, the optimal number of LV intervals is also small. The larger the number of LV intervals, the smaller the number of calibration samples are utilised to calculate the residual quantiles. In highly uncertain scenarios, like the one presented in this case study, a small sample size within an interval does not guarantee representativeness. Increasing the calibration sample size may shift the optimal number of LV intervals to a higher number, and make the PIs more accurate. The low number of optimal LV intervals may also reflect the randomness of the uncertainty. If no discernible pattern emerges from the uncertainty, the factors that contribute to it cannot be systematically modeled with the available data.

It is important to emphasize that the reduced number of optimal LV intervals observed in this case does not imply a need to modify the predictive model. Instead, it reflects limitations in the representativeness of residual quantiles when calibration samples are scarce or uncertainty is dominated by unstructured effects. In such scenarios, increasing discretization resolution does not improve uncertainty calibration and may instead degrade it by producing unreliable quantile estimates.

This case highlights a key limitation of latent-variable-based uncertainty calibration: when uncertainty is dominated by unstructured or random effects, increasing model or discretization complexity does not necessarily improve prediction interval accuracy.

\subsection{Further discussion and development potential}

From the experiments conducted in the case studies, the following observations can be made:

\begin{itemize}
\item The estimated residual quantiles can identify the magnitude of prediction errors using only input-dependent latent projections, enabling spatially meaningful localization of uncertainty.

\item Uncertainty calibration depends on both (i) the number of retained latent variables $H$ (selected by standard predictive criteria) and (ii) the discretization resolution $k$ used to estimate conditional residual quantiles. In cases where the predictive uncertainties are completely at random or they do not depend on the input data (\textbf{X}), the optimal LV interval calibration may converge in a lower number of LVs than at model construction phase. In this case, fewer LVs can be utilised for the uncertainty calibration, while the number of model LVs remain the same.

\item Increasing the number of LV intervals $k$ improves localization up to the point where calibration intervals become sparsely populated; beyond this point, quantile estimates become unstable and prediction intervals can degrade.

\item Coverage-based diagnostics (percentage of test residuals exceeding the nominal PI) and exceedance magnitude provide complementary indicators: the former assesses calibration, while the latter reflects how severe miscalibration is in the tails.

\item In small or well-behaved datasets, conservative coverage can occur for a wide range of discretizations; in this case, the role of discretization is primarily to avoid overly wide (unnecessarily conservative) prediction intervals.

\item Compared to bootstrap baselines, the proposed LV-localized method provides input-adaptive intervals at low computational cost. In particular, na\"{\i}ve (pairs) bootstrap requires refitting the full LV model for each replicate (recomputing $\mathbf{T}$), which can be computationally expensive and may introduce additional variability due to latent-space instability across resamples.
\end{itemize}

Improvements can be made to the current approach. Currently, symmetric PIs are obtained from absolute residual quantiles; this could be extended to asymmetric intervals by using separate quantiles for negative and positive residuals. In addition, discretization could be replaced by a continuous mapping from latent projections to conditional quantiles (e.g., via smoothing or quantile regression in latent space), alleviating sensitivity to interval boundaries.

\section{Conclusion}

The present study proposes a method that embeds an uncertainty calibration procedure in the training of multivariate statistical models, such as PCR, PLS, and their kernelized versions. Previously, these multivariate statistical regression techniques had as an output point estimate as a product, instead of an interval or a distribution. Uncertainty quantification had to be ensured through external or additional studies. The novelty of the research comes from the proposed procedure for estimating PIs for the point predictions, based solely on the input data. 

The data is mapped onto a latent space by default in multivariate statistical models, and those LVs are utilized here for uncertainty calibration as well. The interval on the LV that a new sample is projected on decides the width of the prediction interval for the prediction, based on past experience from the calibration step. 

Along with the proposed procedure, guidelines have been proposed on choosing the number of LVs and the degree of LV discretization. For representativeness, the technique has been tested in scenarios of high- and low-uncertainty, for various spectral regression models, both in remote and laboratory-based scenarios.  

\section*{Acknowledgements}
ZSD, TS, SPR, and HH acknowledge funding from the Research Council of Finland for the Centre of Excellence in Inverse Modelling and Imaging 2018--2025 (decision number 353095) and for the Flagship of Advanced Mathematics for Sensing, Imaging, and Modelling 2024--2031 (decision number 359183). TZ and PT acknowledge funding from US NSF ASCEND Biology Integration Institute (BII) award DBI 2021898 and from NASA Jet Propulsion Laboratory in support of the Surface Biology and Geology (SBG) High Frequency Time Series (SHIFT) through awards 167319 and 1673171. The research effort by JS, AB and OL was carried out at the Jet Propulsion Laboratory, California Institute of Technology, under a contract with the National Aeronautics and Space Administration.

\appendix

\section{Workflow}

\begin{figure}[H]
    \centering
    \includegraphics[width=0.8\linewidth]{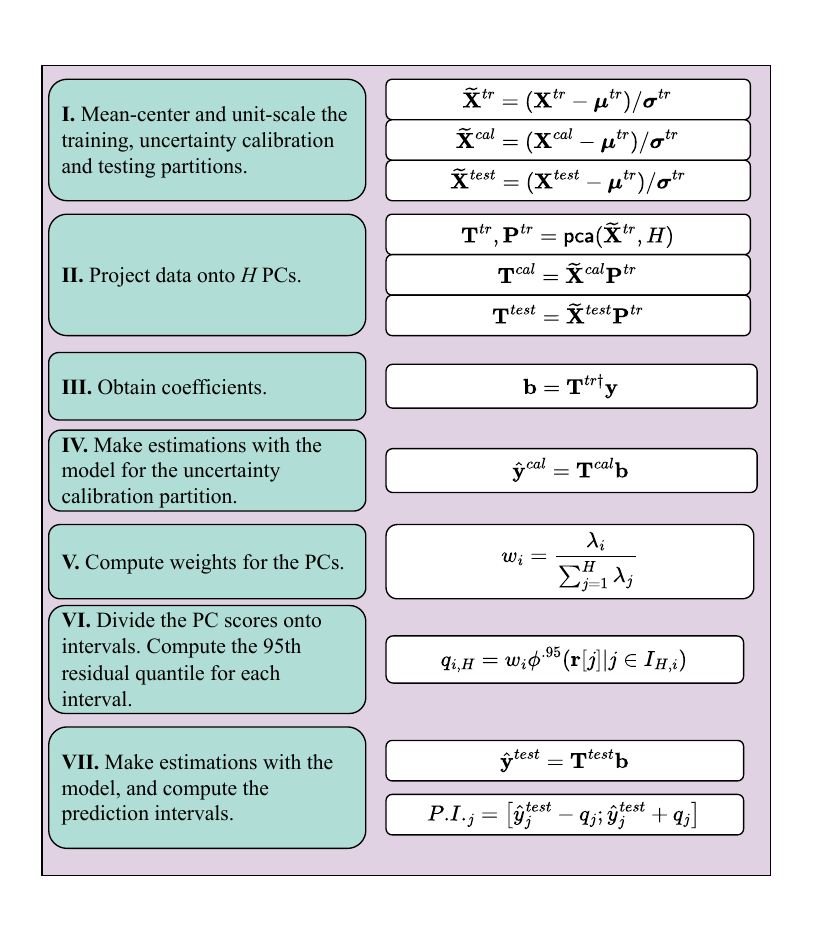}
    \caption{Proposed workflow for uncertainty calibration, for a PCR model. The equations are explained and expanded in Section \ref{sec:maths}.
    \label{fig:steps}}
\end{figure}

\section{Latent-variable regression models and kernel-based extensions}
\label{app:LVmodels}

This appendix summarizes the latent-variable regression models used in the case studies and clarifies how latent projections are obtained in linear and kernel-based settings. These details are provided for completeness and reproducibility, as the uncertainty calibration procedure itself is model-agnostic and depends only on the latent score matrix $\mathbf{T}$.

\subsection{Uncertainty calibration for PLS}

The main difference between PCR and PLS is the rotation of the LVs towards the maximum \textbf{X}-\textbf{Y} covariance in PLS. In addition to the \textbf{X}-side decomposition expressed in Eq.~\ref{eq:PCA}, the \textbf{Y}-side is also decomposed as:
\begin{equation}
    \mathbf{Y} = \mathbf{U}_{\mathrm{PLS}} \mathbf{Q}_{\mathrm{PLS}}^\intercal + \mathbf{F}_{\mathrm{PLS}},
\end{equation}
where $\mathbf{U}_{\mathrm{PLS}}$ is the \textbf{Y}-side score matrix, $\mathbf{Q}_{\mathrm{PLS}}$ is the \textbf{Y}-side loading matrix and $\mathbf{F}_{\mathrm{PLS}}$ is the \textbf{Y}-side residual matrix.

The regression is performed with the coefficients in the original data space:
\begin{equation}
    \mathbf{b}_{\mathrm{PLS}} = \mathbf{W}_{\mathrm{PLS}} (\mathbf{P}_{\mathrm{PLS}}^\intercal \mathbf{W}_{\mathrm{PLS}})^{-1} \mathbf{Q}_{\mathrm{PLS}}^\intercal
\end{equation}
where \textbf{W} are the loadings rotated in the direction of the maximum \textbf{T}-\textbf{Y} covariance.

The absolute residuals ($\mathbf{r}$) for the calibration set are computed for the calibration set 
\begin{equation}
    \textbf{r}^{\mathrm{cal}} = | \mathbf{y}^{\mathrm{cal}} - \widetilde{\mathbf{X}}^{\mathrm{cal}} \mathbf{b}_{\mathrm{PLS}}| \text{.}
\end{equation}

Then, predictions are done with the model
\begin{equation}
    \hat{\mathbf{y}}^{\mathrm{test}} = \widetilde{\mathbf{X}}^{\mathrm{test}} \mathbf{b}_{\mathrm{PLS}} \textbf{.}
\end{equation}

The rest of the procedure is identical as in the PCR example: (i) the X-side loadings ($ \mathbf{T}_{\mathrm{PLS}}^{\mathrm{cal}}$) are discretized in LV intervals for all $H$ LVs included (ii) the residual quantiles are computed for each of the LV intervals (iii) the weights for each of the $H$ selected LVs are calculated based on the explained variance.

\subsection{Uncertainty calibration for K-PCR and K-PLS}

For the kernel variants, data is mapped onto the Reproducing Kernel Hilbert Space, RKHS ($\mathscr{H}$), using a kernel function. For example, using the radial basis function (RBF) kernel we have
\begin{equation}
    \mathsf{k}(\mathbf{x}, \mathbf{x}^*; \boldsymbol{\theta}) 
    = \exp\!\left(- \frac{\lVert \mathbf{x} - \mathbf{x}^* \rVert^2}{2 \sigma^2} \right),
\end{equation}
where $\boldsymbol{\theta} = \{\sigma\}$ denotes the kernel parameter.  
When constructing the kernel (Gram) matrix for a dataset $\mathbf{X} = \{\mathbf{x}_1, \dots, \mathbf{x}_N\}$, the entries are given by
\begin{equation}
    \mathbf{K} = \big[\, \mathsf{k}(\mathbf{x}_i, \mathbf{x}_j; \boldsymbol{\theta}) \,\big]_{i,j=1}^N + \delta \mathbf{I}.
\end{equation}
In the case where one kernel is utilised for the whole dataset $\boldsymbol{\theta} = \{\sigma, \delta\}$, where $\delta$ has the magnitude of the random noise. 

If instead of learning a single kernel width for the whole dataset, we assign one kernel parameter per variable (anisotropic RBF kernel), the kernel function becomes
\begin{equation}
    \mathsf{k}(\mathbf{x}, \mathbf{x}^*; \boldsymbol{\theta}) 
    = \gamma^2 \exp\!\left( - \sum_{i=1}^{m} \frac{(x_i - x_i^*)^2}{2 \sigma_i^2} \right),
\end{equation}
where $\boldsymbol{\theta} = \{\gamma^2, \sigma_1^2, \ldots, \sigma_m^2\}$. 
Here, $\sigma_i^2$ controls the contribution of variable $i$ to the kernel similarity. 
The kernel matrix is then constructed over the sets 
$\mathbf{X}^{\mathrm{tr}}$, $\mathbf{X}^{\mathrm{cal}}$, and $\mathbf{X}^{\mathrm{test}}$, 
with the reference set always taken as the training data ($\mathbf{X}^* = \mathbf{X}^{\mathrm{tr}}$).

Kernel matrices are centered using:
\begin{equation}
   \widetilde{\mathbf{K}}^{\mathrm{tr}} = (\mathbf{I} - \frac{1}{n}\mathbf{1}_n\mathbf{1}_n^\intercal)\mathbf{K}^{\mathrm{tr}}(\mathbf{I} - \frac{1}{n}\mathbf{1}_n\mathbf{1}_n^\intercal)
\end{equation}

where \textbf{I} is the identity matrix and ${\mathbf{1}_n}_{ij} = \frac{1}{n}$ for all $i,j$. The uncertainty calibration ($\mathbf{K}^{\mathrm{cal}}$) and test ($\mathbf{K}^{\mathrm{test}}$) kernel matrices are also centered, with respect to the training set centers \cite{rosipal2001kernel}. The resulting matrices are denoted $\widetilde{\mathbf{K}}^{\mathrm{cal}}$ and $\widetilde{\mathbf{K}}^{\mathrm{test}}$.

The next step after constructing the kernel matrices, is projecting these onto LVs. In K-PCR, the kernel matrix is first decomposed with PCA, as opposed to decomposing the original data matrix as in PCR \cite{hoegaerts2005subset} $\widetilde{\mathbf{K}}^{\mathrm{tr}} = \mathbf{T}_{\mathrm{KPCA}} \mathbf{P}_{\mathrm{KPCA}}^\intercal + \mathbf{E}_{\mathrm{KPCA}}$. Projecting new data onto the LVs for uncertainty calibration purposes is done with:
\begin{equation}
    \mathbf{T}_{\mathrm{KPCR}}^{\mathrm{cal}} = \widetilde{\mathbf{K}}^{\mathrm{cal}} \mathbf{P}_{\mathrm{KPCA}} \textbf{,}
\end{equation}
and the procedure is similar to the traditional PCR from now on, including the division of the LVs onto 
LV intervals and computing the absolute residual quantiles for each LV interval. The regression coefficients are computed as
\begin{equation}
    \mathbf{b}_{\mathrm{KPCR}} = \mathbf{T}_{\mathrm{KPCA}}^{\mathrm{tr}\dagger} \mathbf{y}^{\mathrm{tr}}
\end{equation}
and they are utilised for obtaining the absolute residuals
\begin{equation}
    \textbf{r}^{\mathrm{cal}} = \vert \mathbf{y}^{\mathrm{cal}} - \widetilde{\mathbf{K}}^{\mathrm{cal}} \mathbf{b}_\mathrm{KPCR} \rvert 
\end{equation}
which are utilised as in Eq. \refeq{eq:quantile} - Eq. \refeq{eq:interv}.

In K-PLS, the model is trained using the sequence in Algorithm~\ref{alg:training}. The resulting regression coefficients ($\mathbf{b}_{\mathrm{KPLS}} = \left[b_1, b_2, ..., b_{n}^{\mathrm{tr}}\right]$) are saved for usage in calibration and test predictions. The kernel parameters can be learned via a robust procedure, as presented in Duma \textit{et al.}, 2024 \cite{duma2024kf}.

\begin{algorithm}[H]
\caption{Model training with K-PLS - \textit{kpls}($\widetilde{\mathbf{K}}^{\mathrm{tr}}$, $\mathbf{y}^{\mathrm{tr}}$, $H$) }\label{alg:training}
\textbf{\textit{Data}}: centered kernel matrix ($\widetilde{\mathbf{K}}^{\mathrm{tr}}$), response variable ($\mathbf{y}^{\mathrm{tr}}$), number of LVs ($H$)\\
\textbf{\textit{Output}}: regression coefficients ($\mathbf{b}_{\mathrm{KPLS}}$), predictor-side score matrix ($\mathbf{T}_{\mathrm{KPLS}}^{\mathrm{tr}}$), response-side score matrix ($\mathbf{U}_{\mathrm{KPLS}}^{\mathrm{tr}}$).

\begin{algorithmic}[1]
\For{$i = 1, ..., H$}
    \State $\mathbf{t} \gets \widetilde{\mathbf{K}}^{\mathrm{tr}} \mathbf{y}^{\mathrm{tr}} $ \Comment{Compute K-side score vector for the LV.}
    \State $\mathbf{t} \gets \frac{\mathbf{t}}{\Vert \mathbf{t} \Vert}$ \Comment{Normalize score vector.}
    \State $\mathbf{u} \gets \mathbf{y}^{\mathrm{tr}} (\mathbf{y}^{\mathrm{tr}\intercal} \mathbf{t}) $ \Comment{Compute Y-side score vector for the LV.}
    \State $ \widetilde{\mathbf{K}} \gets \widetilde{\mathbf{K}}^{\mathrm{tr}} - \mathbf{t} (\mathbf{t}^\intercal \widetilde{\mathbf{K}}^{\mathrm{tr}})$  \Comment{Deflate K.}
    \State $ \mathbf{y}^{\mathrm{tr}} \gets \mathbf{y}^{\mathrm{tr}} - \mathbf{t} (\mathbf{t}^\intercal \mathbf{y}^{\mathrm{tr}})$ \Comment{Deflate Y.}
    \State Store $\mathbf{t}, \mathbf{u}$ into $\mathbf{T}^{\mathrm{tr}}_{\mathrm{KPLS}}, \mathbf{U}^{\mathrm{tr}}_{\mathrm{KPLS}}$.
\EndFor

\State 
$\mathbf{b}_{\mathrm{KPLS}}  \gets \mathbf{U}^{\mathrm{tr}}_{\mathrm{KPLS}}\left(\mathbf{T}^{\mathrm{tr}\intercal}_{\mathrm{KPLS}} \widetilde{\mathbf{K}}^{\mathrm{tr}} \mathbf{U}^{\mathrm{tr}}_{\mathrm{KPLS}}\right)^{-1} \mathbf{T}^{\mathrm{tr}\intercal}_{\mathrm{KPLS}} \mathbf{y}^{\mathrm{tr}}$  \Comment{Compute regression coefficients.}

\end{algorithmic}
\end{algorithm}

The uncertainty calibation samples are projected onto the LVs as follows:
\begin{equation}
    \mathbf{T}^{\mathrm{cal}}_{\mathrm{KPLS}} = \widetilde{\mathbf{K}}^{\mathrm{cal}} \mathbf{U}^{\mathrm{tr}}_{\mathrm{KPLS}}\left(\mathbf{T}^{\mathrm{tr}\intercal}_{\mathrm{KPLS}} \widetilde{\mathbf{K}}^{\mathrm{tr}} \mathbf{U}^{\mathrm{tr}}_{\mathrm{KPLS}}\right)^{-1}  \text{,} 
\end{equation}
whereas the absolute residuals are computed as in
\begin{equation}
    \textbf{r}^{\mathrm{cal}} = \vert \mathbf{y}^{\mathrm{cal}} - \widetilde{\mathbf{K}}^{\mathrm{cal}} \mathbf{b}_\mathrm{KPLS} \rvert \text{.}
\end{equation}
The remainder of the procedure follows the methodology in Eq. \refeq{eq:quantile} - \refeq{eq:interv}.

\section{Baseline prediction interval methods and experimental settings}
\label{app:baselinePI}

This appendix provides the exact mathematical definitions of the baseline PI methods used for comparison in Case~1, together with the hyperparameter settings employed in the experiments. The results summarized in Table~2 are based on the formulations given below.

\subsection{Global split conformal prediction intervals}
\label{app:global_conformal}

As a first baseline, we implemented global split conformal prediction for regression. Let
\[
\mathcal{D}_{\mathrm{tr}} = \{(\mathbf{x}_i, y_i)\}_{i=1}^{n_{\mathrm{tr}}}
\]
denote the training set and
\[
\mathcal{D}_{\mathrm{cal}} = \{(\mathbf{x}_j, y_j)\}_{j=1}^{n_{\mathrm{cal}}}
\]
the calibration set. A regression model $\hat{f}(\cdot)$ is fitted using $\mathcal{D}_{\mathrm{tr}}$.

For each calibration sample, the absolute residual (nonconformity score) is computed as
\begin{equation}
    s_j = \left| y_j - \hat{f}(\mathbf{x}_j) \right|, 
    \quad j = 1, \dots, n_{\mathrm{cal}} .
\end{equation}

Let $\widehat{q}_{1-\alpha}$ denote the empirical $(1-\alpha)$ quantile of the set $\{s_j\}_{j=1}^{n_{\mathrm{cal}}}$.  
For a new input $\mathbf{x}_*$ with point prediction $\hat{y}_* = \hat{f}(\mathbf{x}_*)$, the global split conformal prediction interval is given by
\begin{equation}
    \widehat{C}^{\mathrm{global}}_{1-\alpha}(\mathbf{x}_*) =
    \left[
        \hat{y}_* - \widehat{q}_{1-\alpha},
        \;
        \hat{y}_* + \widehat{q}_{1-\alpha}
    \right].
\end{equation}

This method provides marginal coverage guarantees under the assumption of exchangeability between calibration and test samples, but yields constant-width prediction intervals that do not adapt to local uncertainty variations in the input space \cite{vovk2005algorithmic, shafer2008tutorial, lei2018distributionfree}.

\subsection{Residual bootstrap prediction intervals}
\label{app:bootstrap_PI}

As a second baseline, residual bootstrap prediction intervals were constructed using a fixed-design bootstrap scheme, which is commonly applied in chemometric regression.

After fitting the regression model $\hat{f}(\cdot)$, residuals are computed on the calibration set as
\begin{equation}
    \hat{\varepsilon}_j = y_j - \hat{f}(\mathbf{x}_j),
    \quad j = 1, \dots, n_{\mathrm{cal}} .
\end{equation}

For each bootstrap replicate $b = 1, \dots, B$, residuals are resampled with replacement from $\{\hat{\varepsilon}_j\}_{j=1}^{n_{\mathrm{cal}}}$ to form pseudo-responses
\begin{equation}
    y^{(b)}_j = \hat{f}(\mathbf{x}_j) + \hat{\varepsilon}^{(b)}_j .
\end{equation}

Predictions for a new sample $\mathbf{x}_*$ are then obtained as
\begin{equation}
    \hat{y}^{(b)}_* = \hat{f}(\mathbf{x}_*) + \hat{\varepsilon}^{(b)}_* ,
\end{equation}
yielding an empirical distribution of predicted values $\{\hat{y}^{(b)}_*\}_{b=1}^B$.

The $(1-\alpha)$ prediction interval is defined via empirical quantiles as
\begin{equation}
    \widehat{C}^{\mathrm{boot}}_{1-\alpha}(\mathbf{x}_*) =
    \left[
        \phi^{\alpha/2}\!\left(\{\hat{y}^{(b)}_*\}_{b=1}^B\right),
        \;
        \phi^{1-\alpha/2}\!\left(\{\hat{y}^{(b)}_*\}_{b=1}^B\right)
    \right].
\end{equation}

Residual bootstrap prediction intervals rely on the assumption that calibration residuals are representative of future prediction errors. While widely used in chemometrics, they do not provide finite-sample coverage guarantees and can underperform in the presence of heteroscedastic or input-dependent uncertainty \cite{faber1997propagation, martens2000modified, preisner2008uncertainty}.

\subsection{Na\"{\i}ve (pairs) bootstrap prediction intervals}
\label{app:pairs_bootstrap_PI}

As an additional baseline, we implemented the na\"{\i}ve (pairs) bootstrap prediction interval, which is simple and model-agnostic \cite{efron1993bootstrap,davison1997bootstrap}. In contrast to residual bootstrap (Appendix~\ref{app:bootstrap_PI}), the pairs bootstrap resamples observation pairs and refits the full latent-variable model in each replicate (thereby recomputing latent projections).

Let the training set be $\mathcal{D}_{\mathrm{tr}}=\{(\mathbf{x}_i,y_i)\}_{i=1}^{n_{\mathrm{tr}}}$. For each bootstrap replicate $b=1,\ldots,B$, we form a bootstrap sample
\[
\mathcal{D}^{(b)}_{\mathrm{tr}}=\{(\mathbf{x}^{(b)}_i,y^{(b)}_i)\}_{i=1}^{n_{\mathrm{tr}}}
\]
by sampling pairs with replacement from $\mathcal{D}_{\mathrm{tr}}$. A regression model $\hat{f}^{(b)}(\cdot)$ (PCR in Case~1) is then fit on $\mathcal{D}^{(b)}_{\mathrm{tr}}$.

For a new input $\mathbf{x}_*$, the bootstrap predictive distribution is
\begin{equation}
\hat{y}^{(b)}_*=\hat{f}^{(b)}(\mathbf{x}_*), \qquad b=1,\ldots,B.
\end{equation}
The $(1-\alpha)$ na\"{\i}ve bootstrap PI is obtained via percentile endpoints:
\begin{equation}
\widehat{C}^{\mathrm{pairs}}_{1-\alpha}(\mathbf{x}_*)=
\left[
\phi^{\alpha/2}\!\left(\{\hat{y}^{(b)}_*\}_{b=1}^{B}\right),
\;
\phi^{1-\alpha/2}\!\left(\{\hat{y}^{(b)}_*\}_{b=1}^{B}\right)
\right],
\end{equation}
where $\phi^{p}(\cdot)$ denotes the empirical $p$-quantile operator.

\subsection{Hyperparameter settings for Case~1}
\label{app:hyperparameters_case1}

All three uncertainty quantification methods were evaluated using identical data sets and regression models in Case~1. The following hyperparameters were fixed throughout the experiment:

\begin{itemize}
    \item Number of retained latent variables: $H = 3$
    \item Number of latent-variable intervals per LV: $k = 5$
    \item Nominal prediction interval level: $1-\alpha = 0.95$
    \item Number of bootstrap replicates (residual bootstrap): $B = 1000$
    \item Data preprocessing: mean-centering and variance scaling based on the training set
\end{itemize}

Computational time was measured as wall-clock time required to compute prediction intervals for the full test set, excluding model training time.

\bibliography{bibliography}

\begin{thebibliography}{10}
\expandafter\ifx\csname url\endcsname\relax
  \def\url#1{\texttt{#1}}\fi
\expandafter\ifx\csname urlprefix\endcsname\relax\def\urlprefix{URL }\fi
\expandafter\ifx\csname href\endcsname\relax
  \def\href#1#2{#2} \def\path#1{#1}\fi

\bibitem{walker2003defining}
W.~E. Walker, P.~Harremo{\"e}s, J.~Rotmans, J.~P. Van Der~Sluijs, M.~B. Van~Asselt, P.~Janssen, M.~P. Krayer~von Krauss, Defining uncertainty: a conceptual basis for uncertainty management in model-based decision support, Integrated assessment 4~(1) (2003) 5--17.

\bibitem{abdar2021review}
M.~Abdar, F.~Pourpanah, S.~Hussain, D.~Rezazadegan, L.~Liu, M.~Ghavamzadeh, P.~Fieguth, X.~Cao, A.~Khosravi, U.~R. Acharya, et~al., A review of uncertainty quantification in deep learning: Techniques, applications and challenges, Information fusion 76 (2021) 243--297.

\bibitem{jorgensen2023extensible}
S.~Jorgensen, J.~Holodnak, J.~Dempsey, K.~de~Souza, A.~Raghunath, V.~Rivet, N.~DeMoes, A.~Alejos, A.~Wollaber, Extensible machine learning for encrypted network traffic application labeling via uncertainty quantification, IEEE Transactions on Artificial Intelligence 5~(1) (2023) 420--433.

\bibitem{passos2022tutorial}
D.~Passos, P.~Mishra, A tutorial on automatic hyperparameter tuning of deep spectral modelling for regression and classification tasks, Chemometrics and Intelligent Laboratory Systems 223 (2022) 104520.

\bibitem{gromski2015tutorial}
P.~S. Gromski, H.~Muhamadali, D.~I. Ellis, Y.~Xu, E.~Correa, M.~L. Turner, R.~Goodacre, A tutorial review: Metabolomics and partial least squares-discriminant analysis--a marriage of convenience or a shotgun wedding, Analytica chimica acta 879 (2015) 10--23.

\bibitem{wang2022recent}
H.-P. Wang, P.~Chen, J.-W. Dai, D.~Liu, J.-Y. Li, Y.-P. Xu, X.-L. Chu, Recent advances of chemometric calibration methods in modern spectroscopy: Algorithms, strategy, and related issues, TrAC Trends in Analytical Chemistry 153 (2022) 116648.

\bibitem{kamoske2021leaf}
A.~G. Kamoske, K.~M. Dahlin, S.~P. Serbin, S.~C. Stark, Leaf traits and canopy structure together explain canopy functional diversity: an airborne remote sensing approach, Ecological Applications 31~(2) (2021) e02230.

\bibitem{ji2022review}
C.~Ji, W.~Sun, A review on data-driven process monitoring methods: Characterization and mining of industrial data, Processes 10~(2) (2022) 335.

\bibitem{zhang2009comparison}
L.~Zhang, S.~Garcia-Munoz, A comparison of different methods to estimate prediction uncertainty using partial least squares (pls): a practitioner's perspective, Chemometrics and intelligent laboratory systems 97~(2) (2009) 152--158.

\bibitem{faber1997propagation}
K.~Faber, B.~R. Kowalski, Propagation of measurement errors for the validation of predictions obtained by principal component regression and partial least squares, Journal of Chemometrics: A Journal of the Chemometrics Society 11~(3) (1997) 181--238.

\bibitem{martens2000modified}
H.~Martens, M.~Martens, Modified jack-knife estimation of parameter uncertainty in bilinear modelling by partial least squares regression (plsr), Food quality and preference 11~(1-2) (2000) 5--16.

\bibitem{wentzell2015errors}
P.~D. Wentzell, The errors of my ways: Maximum likelihood pca seventeen years after bruce, in: 40 Years of Chemometrics--From Bruce Kowalski to the Future, ACS Publications, 2015, pp. 31--64.

\bibitem{babamoradi2013bootstrap}
H.~Babamoradi, F.~van~den Berg, {\AA}.~Rinnan, Bootstrap based confidence limits in principal component analysis—a case study, Chemometrics and Intelligent Laboratory Systems 120 (2013) 97--105.

\bibitem{preisner2008uncertainty}
O.~Preisner, J.~A. Lopes, J.~C. Menezes, Uncertainty assessment in ft-ir spectroscopy based bacteria classification models, Chemometrics and Intelligent Laboratory Systems 94~(1) (2008) 33--42.

\bibitem{de2013discrimination}
M.~R. de~Almeida, D.~N. Correa, W.~F. Rocha, F.~J. Scafi, R.~J. Poppi, Discrimination between authentic and counterfeit banknotes using raman spectroscopy and pls-da with uncertainty estimation, Microchemical Journal 109 (2013) 170--177.

\bibitem{rocha2018classification}
W.~F. d.~C. Rocha, D.~A. Sheen, D.~W. Bearden, Classification of samples from nmr-based metabolomics using principal components analysis and partial least squares with uncertainty estimation, Analytical and bioanalytical chemistry 410 (2018) 6305--6319.

\bibitem{gibbs2021adaptive}
I.~Gibbs, E.~Candes, Adaptive conformal inference under distribution shift, Advances in Neural Information Processing Systems 34 (2021) 1660--1672.

\bibitem{einbinder2022training}
B.-S. Einbinder, Y.~Romano, M.~Sesia, Y.~Zhou, Training uncertainty-aware classifiers with conformalized deep learning, Advances in Neural Information Processing Systems 35 (2022) 22380--22395.

\bibitem{lei2018distribution}
J.~Lei, M.~G’Sell, A.~Rinaldo, R.~J. Tibshirani, L.~Wasserman, Distribution-free predictive inference for regression, Journal of the American Statistical Association 113~(523) (2018) 1094--1111.

\bibitem{montgomery2012regression}
D.~C. Montgomery, E.~A. Peck, G.~G. Vining, \href{https://books.google.com/books?isbn=9780470542811}{Introduction to Linear Regression Analysis}, 5th Edition, Wiley, 2012.
\newline\urlprefix\url{https://books.google.com/books?isbn=9780470542811}

\bibitem{martens1989multivariate}
H.~Martens, T.~N{\ae}s, \href{https://books.google.com/books?isbn=9780471930471}{Multivariate Calibration}, John Wiley \& Sons, 1989.
\newline\urlprefix\url{https://books.google.com/books?isbn=9780471930471}

\bibitem{faber2002uncertainty}
N.~M. Faber, \href{https://doi.org/10.1016/S0169-7439(02)00102-8}{Uncertainty estimation for multivariate regression coefficients}, Chemometrics and Intelligent Laboratory Systems 64~(2) (2002) 169--179.
\newblock \href {https://doi.org/10.1016/S0169-7439(02)00102-8} {\path{doi:10.1016/S0169-7439(02)00102-8}}.
\newline\urlprefix\url{https://doi.org/10.1016/S0169-7439(02)00102-8}

\bibitem{efron1993bootstrap}
B.~Efron, R.~J. Tibshirani, \href{https://books.google.com/books?isbn=9780412042317}{An Introduction to the Bootstrap}, Chapman \& Hall/CRC, 1993.
\newline\urlprefix\url{https://books.google.com/books?isbn=9780412042317}

\bibitem{davison1997bootstrap}
A.~C. Davison, D.~V. Hinkley, \href{https://www.cambridge.org/core/books/bootstrap-methods-and-their-application/ED2FD043579F27952363566DC09CBD6A}{Bootstrap Methods and Their Application}, Cambridge University Press, 1997.
\newline\urlprefix\url{https://www.cambridge.org/core/books/bootstrap-methods-and-their-application/ED2FD043579F27952363566DC09CBD6A}

\bibitem{wu1986jackknife}
C.~F.~J. Wu, \href{https://doi.org/10.1214/aos/1176350142}{Jackknife, bootstrap and other resampling methods in regression analysis}, The Annals of Statistics 14~(4) (1986) 1261--1295.
\newblock \href {https://doi.org/10.1214/aos/1176350142} {\path{doi:10.1214/aos/1176350142}}.
\newline\urlprefix\url{https://doi.org/10.1214/aos/1176350142}

\bibitem{shafer2008tutorial}
G.~Shafer, V.~Vovk, A tutorial on conformal prediction, Journal of Machine Learning Research 9 (2008) 371--421.

\bibitem{lei2018distributionfree}
J.~Lei, M.~G'Sell, A.~Rinaldo, R.~J. Tibshirani, L.~Wasserman, Distribution-free predictive inference for regression, Journal of the American Statistical Association 113~(523) (2018) 1094--1111.
\newblock \href {https://doi.org/10.1080/01621459.2017.1307116} {\path{doi:10.1080/01621459.2017.1307116}}.

\bibitem{vovk2005algorithmic}
V.~Vovk, A.~Gammerman, G.~Shafer, Algorithmic Learning in a Random World, Springer, New York, 2005.
\newblock \href {https://doi.org/10.1007/b106715} {\path{doi:10.1007/b106715}}.

\bibitem{romano2019conformalized}
Y.~Romano, E.~Patterson, E.~J. Cand{\`e}s, Conformalized quantile regression, Advances in Neural Information Processing Systems 32 (2019).

\bibitem{angelopoulos2021gentle}
A.~N. Angelopoulos, S.~Bates, A gentle introduction to conformal prediction and distribution-free uncertainty quantification, Foundations and Trends in Machine Learning 16~(4) (2023) 494--591.
\newblock \href {https://doi.org/10.1561/2200000101} {\path{doi:10.1561/2200000101}}.

\bibitem{kettaneh2005pca}
N.~Kettaneh, A.~Berglund, S.~Wold, Pca and pls with very large data sets, Computational Statistics \& Data Analysis 48~(1) (2005) 69--85.

\bibitem{iupac_orangebook_2014}
D.~T. Burns, K.~Danzer (Eds.), Compendium of Analytical Nomenclature, 3rd Edition, International Union of Pure and Applied Chemistry, 2014, the Orange Book.

\bibitem{wold2001pls}
S.~Wold, M.~Sj{\"o}str{\"o}m, L.~Eriksson, Pls-regression: a basic tool of chemometrics, Chemometrics and Intelligent Laboratory Systems 58~(2) (2001) 109--130.
\newblock \href {https://doi.org/10.1016/S0169-7439(01)00155-1} {\path{doi:10.1016/S0169-7439(01)00155-1}}.

\bibitem{rosipal2001kernel}
R.~Rosipal, L.~J. Trejo, Kernel partial least squares regression in reproducing kernel hilbert space, Journal of machine learning research 2~(Dec) (2001) 97--123.

\bibitem{hoegaerts2005subset}
L.~Hoegaerts, J.~A. Suykens, J.~Vandewalle, B.~De~Moor, Subset based least squares subspace regression in rkhs, Neurocomputing 63 (2005) 293--323.

\bibitem{riese2018hyperspectral}
F.~M. Riese, S.~Keller, Hyperspectral benchmark dataset on soil moisture, in: Proceedings of the 2018 IEEE International Geoscience and Remote Sensing Symposium (IGARSS), Valencia, Spain, 2018, pp. 22--27.

\bibitem{chadwick2025unlocking}
K.~D. Chadwick, F.~Davis, K.~R. Miner, R.~Pavlick, M.~Reynolds, P.~A. Townsend, P.~G. Brodrick, C.~Ade, J.~Allen, L.~Anderegg, et~al., Unlocking ecological insights from sub-seasonal visible-to-shortwave infrared imaging spectroscopy: The shift campaign, Ecosphere 16~(3) (2025) e70194.

\bibitem{brodrick2023shift}
P.~Brodrick, R.~Pavlick, M.~Bernas, J.~Chapman, R.~Eckert, M.~Helmlinger, M.~Hess-Flores, L.~Rios, F.~Schneider, M.~Smyth, et~al., Shift: Aviris-ng full-resolution true color images. ornl daac, oak ridge, tennessee, usa (2023).

\bibitem{zheng2024shift}
T.~ZHENG, N.~QUEALLY, K.~CHADWICK, J.~CRYER, P.~REIM, C.~VILLANUEVA-WEEKS, P.~TOWNSEND, M.~BERG, Z.~BREUER, N.~BURKARD, et~al., Shift: Laboratory foliar chemical analysis results for field samples, ca, 2022, ORNL DAAC (2024).

\bibitem{duma2024kf}
Z.-S. Duma, J.~Susiluoto, O.~Lamminp{\"a}{\"a}, T.~Sihvonen, S.-P. Reinikainen, H.~Haario, Kf-pls: Optimizing kernel partial least-squares (k-pls) with kernel flows, Chemometrics and Intelligent Laboratory Systems 254 (2024) 105238.

\end{thebibliography}

\end{document}